\documentclass{article}

\usepackage[a4paper, top=2cm, bottom=2cm, left=3cm, right=3cm]{geometry}

\usepackage[numbers,super,sort&compress]{natbib}
\usepackage[utf8]{inputenc}
\usepackage[T1]{fontenc}
\usepackage{graphicx}%
\usepackage{multirow}%
\usepackage{amsmath,amssymb,amsfonts}%
\usepackage{amsthm}%
\usepackage{stmaryrd}
\usepackage{mathrsfs}%
\usepackage{hyperref}
\usepackage{dsfont}
\usepackage[title]{appendix}%
\usepackage{xcolor}%
\usepackage{textcomp}%
\usepackage{manyfoot}%
\usepackage{booktabs}%
\usepackage{bm}
\usepackage[inkscapelatex=false]{svg}
\usepackage{tikz}
\usepackage{tikzscale}
\usepackage{booktabs}
\usepackage{pifont}
\usepackage{textcomp}

\usepackage{textcomp}
\usepackage{float}
\usepackage{cleveref}
\usepackage{mathtools}
\usepackage{comment}
\usepackage{booktabs}
\usepackage[acronym]{glossaries}
\usepackage{caption}
\usepackage{authblk}

\glsdisablehyper
\newacronym{massspec}{LC-MS}{\textit{Liquid Chromatography-Mass Spectrometry}}
\newacronym{ivt}{IVT}{\textit{in vitro} transcription}
\newacronym{igv}{IGV}{Integrative Genomics Viewer}
\newacronym{ont}{ONT}{Oxford Nanopore Technologies}
\newacronym[
  plural={NN scores}
]{nns}{NN score}{nearest neighbor anomaly score}

\newacronym{bh}{BH}{Benjamini–Hochberg}
\newacronym{denv}{DENV}{dengue virus}

\Crefformat{figure}{#2Fig.~#1#3}
\Crefmultiformat{figure}{Figs.~#2#1#3}{ and~#2#1#3}{, #2#1#3}{ and~#2#1#3}

\usepackage{algorithm}%
\usepackage{algorithmicx}%
\usepackage{algpseudocode}%
\usepackage{listings}%

\newcommand{\poly}[1]{#1}
\newcommand{\score}[2]{s(#1;#2)}
\newcommand{\pathx}[0]{X}

\newcommand{\Signature}[1]{\phi_{#1}}

\usepackage{cleveref}
\usepackage{mathtools}
\usepackage{subcaption} 
\usepackage{amsmath}
\Crefformat{figure}{#2Fig.~#1#3}
\Crefmultiformat{figure}{Figs.~#2#1#3}{ and~#2#1#3}{, #2#1#3}{ and~#2#1#3}
\Crefformat{section}{#2Sec.~#1#3}
\Crefformat{algorithm}{#2Alg.~#1#3}
\Crefformat{theorem}{#2Thm.~#1#3}
\Crefformat{definition}{#2Def.~#1#3}

\title{
Path Signatures Enable Model-Free Mapping \\ of RNA Modifications}

\author[1,*,\dag]{Maud Lemercier}
\author[2,*]{Paola Arrubarrena}
\author[3,*]{Salvatore Di Giorgio}

\author[4]{Julia Brettschneider}
\author[2]{Thomas Cass}
\author[3]{Valerie Griesche}
\author[6,7]{Isabel S. Naarmann-de Vries}
\author[4]{Anastasia Papavasiliou}
\author[5]{Alessia Ruggieri}
\author[3]{Irem Tellioglu}
\author[5]{Chia Ching Wu}
\author[3,\dag]{F. Nina Papavasiliou}
\author[1,\dag]{Terry Lyons}

\affil[1]{Mathematical Institute, University of Oxford, Oxford, UK.}
\affil[2]{Department of Mathematics, Imperial College London, London, UK.}
\affil[3]{Division of Immune Diversity, German Cancer Research Center, Heidelberg 69120, Germany.}
\affil[4]{Department of Statistics, University of Warwick, Coventry, UK.}
\affil[5]{Department of Infectious Diseases, Molecular Virology, Center for Integrative Infectious Disease Research, Heidelberg University, Medical Faculty Heidelberg, 69120 Heidelberg, Germany.}
\affil[6]{Klaus Tschira Institute for Integrative Computational Cardiology, Heidelberg University, Medical Faculty Heidelberg, 69120 Heidelberg, Germany.}
\affil[7]{German Centre for Cardiovascular Research (DZHK)-Partner Site Heidelberg/Mannheim, 69120 Heidelberg, Germany.}

\date{}

\begin{document}

\maketitle

\begingroup
\renewcommand{\thefootnote}{\fnsymbol{footnote}}

\footnotetext[1]{These authors contributed independently and are designated joint first authors.}

\footnotetext[2]{Corresponding authors: M.L. (\texttt{m.lemercier@logosresearch.ai}) and T.L. (\texttt{terry.lyons@maths.ox.ac.uk}; \texttt{t.lyons@imperial.ac.uk}) for models and software; F.N.P. (\texttt{n.papavasiliou@dkfz-heidelberg.de}) for biological interpretation, datasets and reagents.}

\endgroup

\abstract{Detecting chemical modifications on RNA molecules remains a key challenge in epitranscriptomics. Traditional reverse transcription-based sequencing methods introduce enzyme- and sequence-dependent biases and fragment RNA molecules, confounding the accurate mapping of modifications across the transcriptome. Nanopore direct RNA sequencing offers a powerful alternative by preserving native RNA molecules, enabling the detection of modifications at single-molecule resolution. However, current computational tools can identify only a limited subset of modification types within well-characterized sequence contexts for which ample training data exists. Here, we introduce a model-free computational method that reframes modification detection as an anomaly detection problem, requiring only canonical (unmodified) RNA reads without any other annotated data. For each nanopore read, our approach extracts robust, modification-sensitive features from the raw ionic current signal at a site using the signature transform, then computes an anomaly score by comparing the resulting feature vector to its nearest neighbors in an unmodified reference dataset. We convert anomaly scores into statistical p-values to enable anomaly detection at both individual read and site levels. Validation on densely-modified \textit{E. coli} rRNA demonstrates that our approach detects known sites harboring diverse modification types, without prior training on these modifications. We further applyied this framework to \acrfull{denv} transcripts and mammalian mRNAs. For DENV sfRNA, it led to revealing a novel 2'-O-methylated site, which we validate orthogonally by qRT-PCR assays. These results demonstrate that our model-free approach operates robustly across different types of RNAs and datasets generated with different nanopore sequencing chemistries.}

\section{Introduction}

\begin{figure}[!t]
    \centering
   \begin{subfigure}{0.20\textwidth}
\includegraphics[width=1.\linewidth]{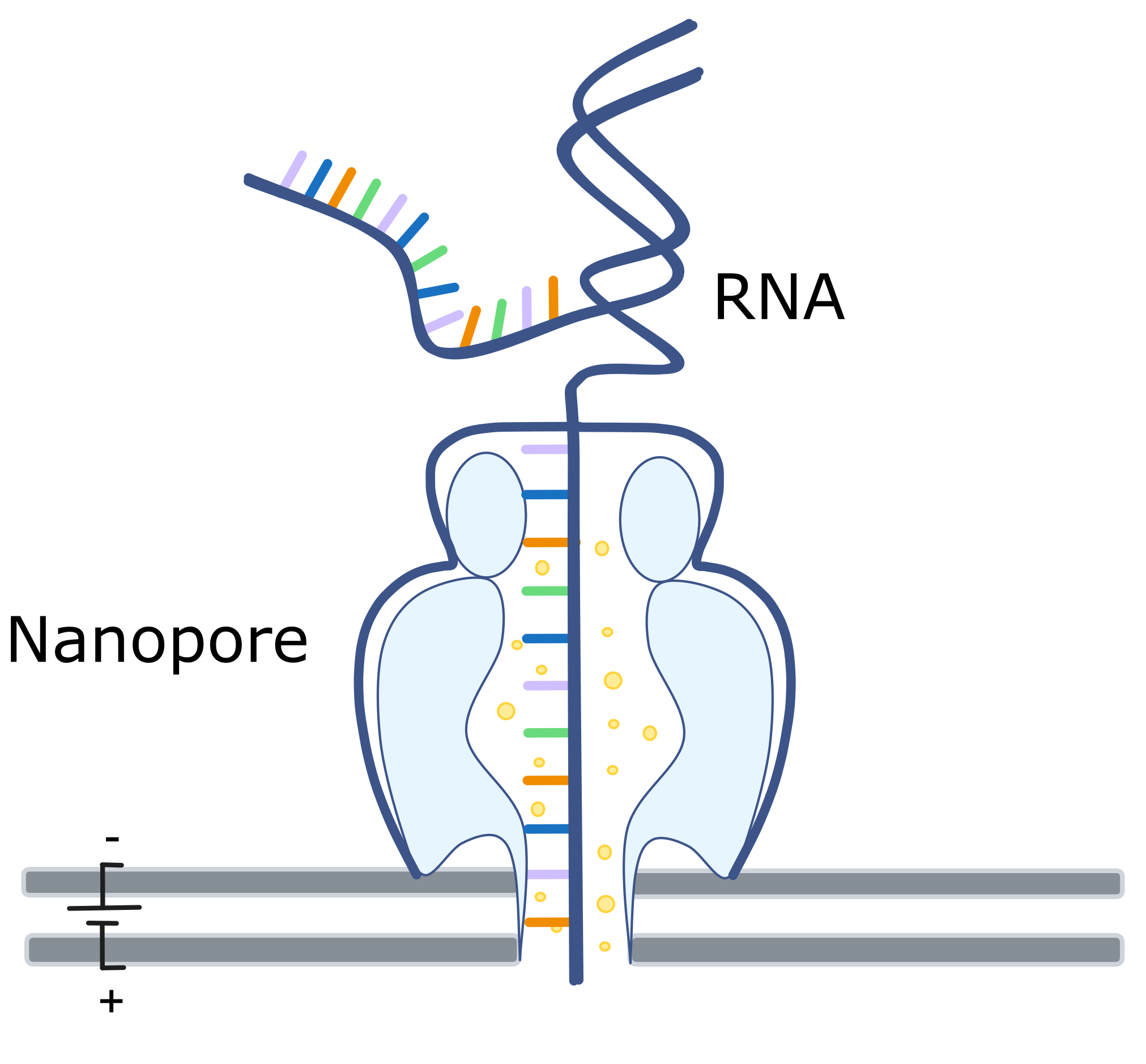}
\caption{}
\label{fig:nanopore}
    \end{subfigure}
     \begin{subfigure}{0.50\textwidth}
\includegraphics[width=1\linewidth]{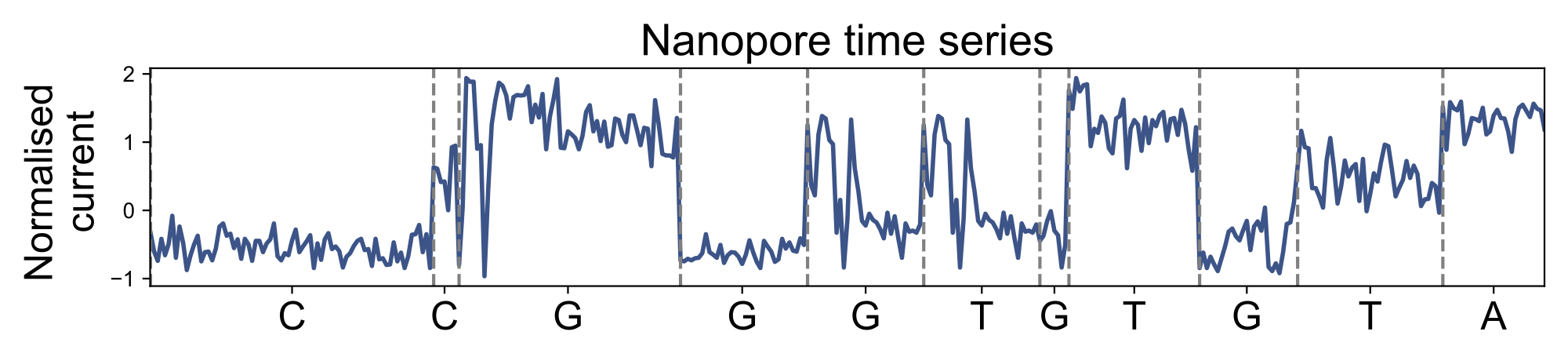}
\caption{}
\label{fig:basecalling_alignment}
    \end{subfigure}    
\\
    \begin{subfigure}[b]{0.23\textwidth}
\includegraphics[width=1.\linewidth]{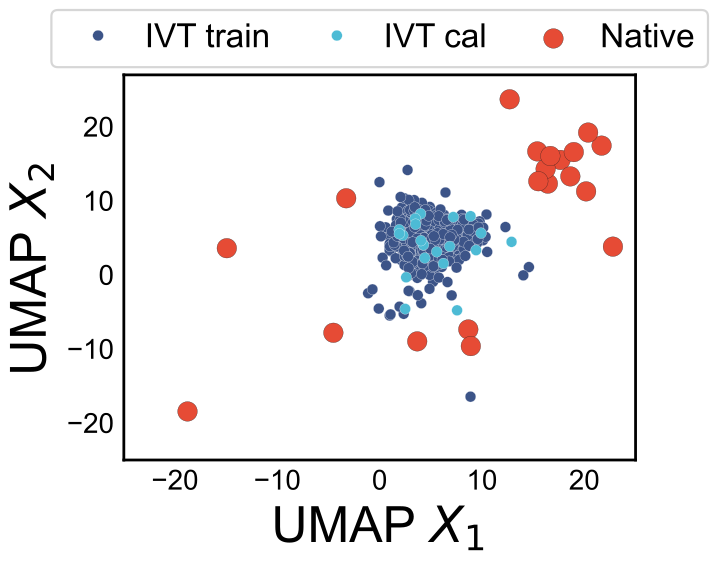}
\caption{}
\label{fig:signatures_umap}
    \end{subfigure}    
     \begin{subfigure}[b]{0.23\textwidth}
\includegraphics[width=1\linewidth]{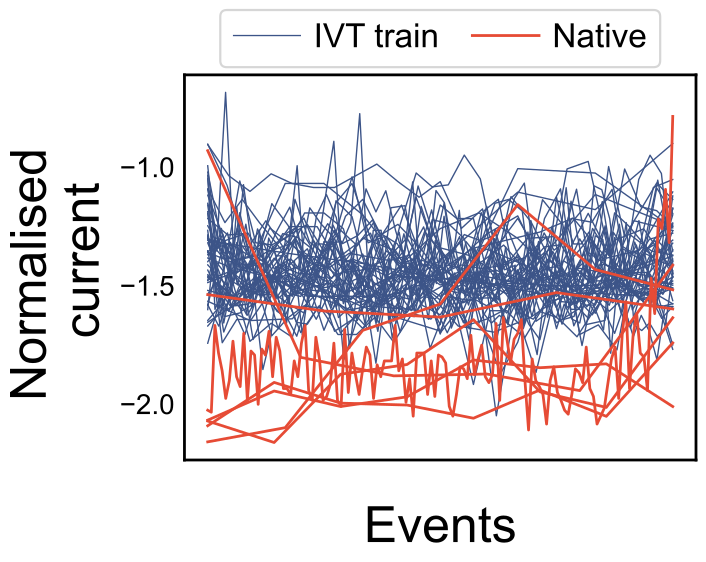}
\caption{}
\label{fig:nanopore_signals}
    \end{subfigure}
    \begin{subfigure}[b]{0.23\textwidth}
\includegraphics[width=1\linewidth]{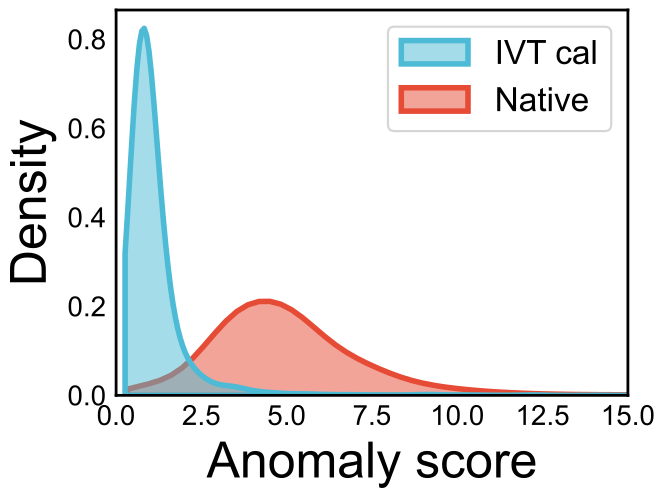}
\caption{}
\label{fig:conformance_score}
    \end{subfigure}
    
\hfill 

     \begin{subfigure}{0.20\textwidth}
\includegraphics[width=1\linewidth]{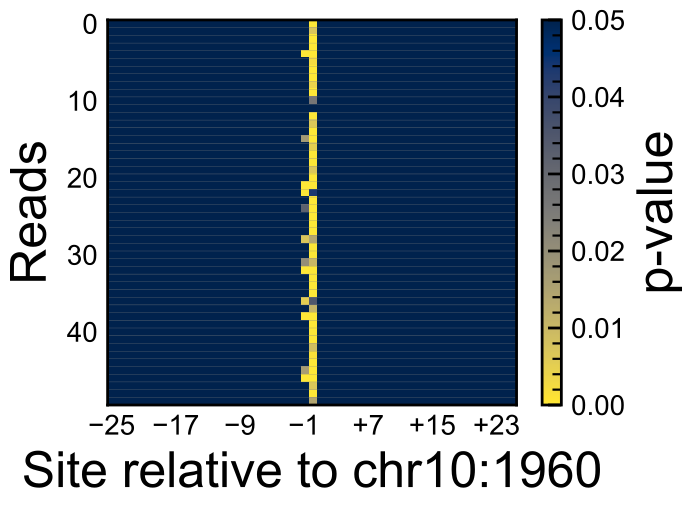}
\caption{}
\label{fig:anomaly_map}
    \end{subfigure}
     \begin{subfigure}{0.50\textwidth}
\raisebox{4mm}{\includegraphics[width=1\linewidth]{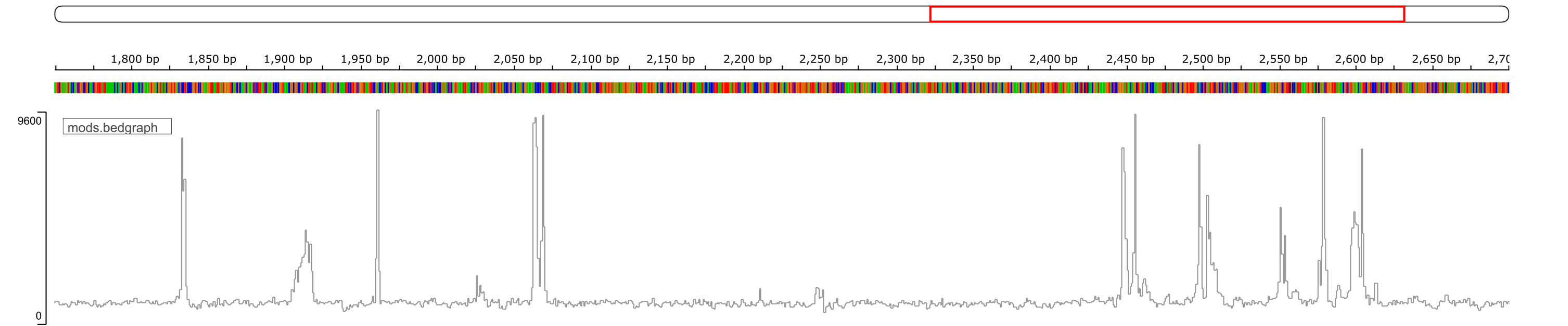}}
\caption{}
\label{fig:Bedgraph}
\end{subfigure}
\\
\begin{subfigure}{0.65\textwidth}
  \centering
  \ttfamily
   \footnotesize
  \begin{tabular}{@{} l r r r r r r@{}}
    \toprule
    \textbf{chrom} & \textbf{start} & \textbf{end} 
       & \textbf{n\_test}  & \textbf{n\_cal} & \textbf{n\_anom@0.1} & \textbf{Fisher\_pval}\\
       \midrule 
       chr10 & 51484 & 514855 & 3512 & 1590 &  976 & 2.86e-52  \\
       chr10 & 23479 & 23480 & 16 & 2000 & 0 & 0.99 \\ 
    \bottomrule
  \end{tabular}
\caption{}
\label{fig:bedmethyl}
\end{subfigure}
    \caption{\textbf{Schematic of the workflow.} (a) Schematic of an RNA molecule threading through a nanopore. (b) Example of an ionic current time series from a nanopore read aligned to a reference sequence. (c) 2D visualization of the signature embeddings of IVT and modified current stretches using UMAP. (d) Corresponding ionic current stretches. (e) Densities estimated with the corresponding anomaly scores. (f) Visualization of p-values per-site and per-read. (g) IGV visualization of a BedGraph file recording per-site modification inferences, such as the frequency of anomalous reads at a predefined significance threshold and the site-level p-value combining the p-values across reads at a site (h) BED file format. \texttt{n\_anom@0.01}: number of anomalies detected by thresholding the conformal p-values at $0.1$; \texttt{Fisher\_pval}: Fisher's combination test p-value (testing the hypothesis that no read at a site is anomalous), where the underlying test statistic combines the \texttt{n\_test} conformal p-values, after adjustment.}
\end{figure}

\hspace{0pt}\\Detecting and mapping RNA modifications is a fundamental challenge in molecular biology, with profound implications for understanding gene regulation, cellular function, and disease mechanisms \cite{boccalettoMODOMICSDatabaseRNA2018, saletoreBirthEpitranscriptomeDeciphering2012}. These modifications play pivotal roles in diverse biological processes, influencing RNA stability, localization, and interactions \cite{songModificationRRNAQuality2002}. Despite their importance, the precise distribution and functional roles of many chemical modifications remain poorly understood. Addressing this knowledge gap is critical, as dysregulation of modification pathways has been implicated in a wide range of disorders, including cancer, neurodegeneration, and metabolic diseases \cite{hsuEpitranscriptomicInfluencesDevelopment2017}. Recent advances in biomolecular sensing technologies, such as nanopore-based sequencing, coupled with computational methods, have catalyzed significant progress in epitranscriptomics. These technologies enable direct and high-throughput detection of modifications, bypassing the need for labor intensive chemical derivatization or antibody-based enrichment \cite{garaldeHighlyParallelDirect2018,liuAccurateDetectionM6A2019a, lorenzDirectRNASequencing2020a}. Nanopore sequencing, in particular, has emerged as a transformative approach, offering single-molecule resolution and the ability to detect modifications based on alterations in ionic current signals as nucleic acids pass through the pore \cite{garaldeHighlyParallelDirect2018, smithReadingCanonicalModified2019}. This provides unprecedented opportunities for transcriptome-wide mapping of chemical modifications.

Specifically, nanopore flow cells transduce polymeric molecules into electrical signals. Neural networks trained on extensive datasets (pairing the data strings of A, C, G, T (or U), with their corresponding ionic current measurements) are used to decode these signals directly, a process known as \emph{basecalling} (\Cref{fig:nanopore,fig:basecalling_alignment}). Beyond sequencing, the variations in nanopore electrical signals offer a promising avenue for detecting chemically modified nucleotides. In the task of \emph{modification calling}, the objective is to determine, based on a nanopore signal, whether an RNA (or DNA) molecule comprises solely canonical nucleotides or also harbors chemical modifications \cite{acera2023concepts}.  

However, current computational methods for analyzing nanopore sequencing data face important limitations. On the one hand, most existing tools are tailored to detect only specific modifications, such as N6-methyladenosine (m6A) or 5-methylcytosine (m5C). Recently, \acrfull{ont} has expanded this scope by adding support for 2'-O-methylations (2OmeA, 2OmeC,  2OmeU and 2OmeG), inosine (I), and pseudouridine ($\Psi$) in their proprietary tool Dorado. While these modification-aware basecallers are useful when a particular known modification is sought, their development relies on collecting large, modification-annotated training datasets \cite{begikDecodingRibosomalRNA2020, konoNanoporeSequencingReview2019, jonkhoutRNAModificationLandscape2017, novoaChartingUnknownEpitranscriptome2017, esfahani2025evaluation}. Each base requires its own machine learning classifier (or bespoke model weights), forcing users to run multiple, modification-specific predictors. This not only incurs substantial computational overhead but also precludes the discovery of novel modification types beyond the scope of the pre-trained models, which need to be retrained whenever ONT releases a new pore version or sequencing kit. Furthermore, any out-of-distribution signal (potentially caused by a truly novel modification) can go undetected or be mislabelled. 

On the other hand, unsupervised clustering approaches \cite{ueda2023rna,leger2021rna,pratanwanich2021identification,riquelme2025direct} have shown promise for detecting modifications without requiring large annotated datasets. However, most current implementations fall short of achieving single-molecule resolution across a diverse set of modification types. Apart from Nanodoc2 \cite{ueda2023rna}, a deep one-class classifier trained on modification-free data, which needs retraining when the sequencing chemistry changes, these tools generally collapse the nanopore time series to only one or two summary features, potentially limiting their sensitivity to subtle modification-induced signal changes. 

To address these gaps, we introduce a semi-supervised computational framework that detects unexpected patterns in the time-series of electrical signals that come through the nanopore as it reads a single molecule. Our method builds a baseline corpus of unmodified nanopore signals, generated using modification-free transcripts from \acrfull{ivt} of cDNA and uses the signature transform \cite{chevyrev2016primer, morrill2020generalised, fermanian2023new, lyons2025signature} to extract rich statistical and temporal features from ionic current time series. By comparing each read's feature vector to its modification-free counterparts, this framework reliably flags anomalous sites across a comprehensive set of chemical modifications. Although it does not output the type of modification, the ranked anomaly scores it produces enable rapid prioritization of candidate sites for downstream classification or orthogonal experimental validation.
Implemented as a scalable pipeline for high-throughput nanopore sequencing data, our tool produces a \emph{transcriptome-wide BED summary file} (\Cref{fig:bedmethyl}) and \emph{per-read per-site anomaly maps} (\Cref{fig:anomaly_map}) for any smaller region of interest. Together, these outputs make it straightforward to pinpoint sites that require downstream validation and identification of modification type. Furthermore, our model-free framework is applicable to virtually all RNA species and sequencing chemistries. We analyzed molecules of different kinds, ranging from bacterial ribosomal RNAs (rRNAs), dengue virus subgenomic flaviviral RNA (sfRNA) to mRNAs derived from mouse cell lines sequenced with RNA002 and RNA004 nanopore sequencing chemistries.

\section{Results} 
\subsection{Anomaly detection in nanopore sequencing signals}

We quantify the degree of novelty of nanopore current time series by comparing them to time series from canonical polymers. To build this reference dataset, we sequence RNA that is transcribed \emph{in vitro} from cDNA using only canonical nucleotides, which in turn is reverse transcribed from the cell's native RNA, thus preserving the sequence context while removing RNA modifications. For each read in an experimental sample, we partition its current trace into successive signal segments using the nanopore signal alignment tool Uncalled4 \cite{kovakaUncalled4ImprovesNanopore2024}, and apply the signature transform to each segment to obtain fixed-length feature vectors (\Cref{fig:signatures_umap}) that faithfully capture temporal dynamics (\Cref{fig:nanopore_signals}). We then score each vector by its nearest neighbor Mahalanobis distance \cite{shao2020dimensionless, arrubarrena2024novelty} within the IVT feature vectors. Here, higher \glspl{nns} indicate a greater deviation from canonical behavior, suggesting the presence of anomalies likely induced by RNA chemical modifications, or, less commonly, by unfiltered genomic variants or rare sequencing or alignment artefacts. To make these scores comparable across different sites, and set threshold values, we calibrate them against an independent set of IVT reads, turning each raw \gls{nns} (for each read-site pair) into a so-called conformal p-value \cite{bates2023testing}. Starting from nanopore electrical signals aligned to a reference sequence, our pipeline thus assigns to each site along every read an anomaly score together with a p-value, which can be visualized as an anomaly map (\Cref{fig:anomaly_map}). All per-site summary statistics—including coverage and anomaly counts—are reported in a BED file (\Cref{fig:bedmethyl}), which can be readily converted into a BedGraph for genome browser visualization (\Cref{fig:Bedgraph}). Statistical inference is supported at two complementary levels: significant site–read pairs can be flagged directly, or evidence can be aggregated across reads into site-level p-values; in both cases, thresholding and multiple testing correction highlight putative RNA modifications. Our signature-based anomaly detection framework includes three key advances:\\ 

\noindent\textbf{Statistically calibrated anomaly scores.} Unlike clustering approaches (e.g. Nanocompore) or deep classifiers (e.g. Dorado or m6ABasecaller) that output modification probabilities that may be uncalibrated, difficult to interpret and often leading to high false positive rates \cite{diensthuber2025systematic}, we convert our anomaly scores into statistically valid p-values, which enables principled false discovery rate control across thousands of reads and sites tested simultaneously. Individual p-values can be visualized as heatmaps using custom scripts or directly in genome browsers such as \acrfull{igv}. For site-level inference, we aggregate p-values across reads using established statistical combination methods to obtain global p-values. \\
    
\noindent\textbf{Transcriptome-wide detection of anomalous sites.} 
By thresholding p-values and counting anomalous reads at each genomic position we obtain an \emph{anomaly rate}. Although this quantity is distinct from modification stoichiometry (the proportion of molecules carrying a modification at a given site), it provides a useful measure for prioritizing loci for further analysis. High-confidence sites can then be passed to discriminative classifiers or orthogonal validation assays to determine the modification types, thereby reducing both computational load and experimental effort. All per-site statistics, together with coverage information, are reported in BED format, facilitating interoperability with genome browsers and downstream analysis pipelines. This representation also enables direct comparison between conditions: for example, differential anomaly rates between samples can be estimated using a beta-binomial proportion test that accounts for calibration and sample size, a strategy that is particularly informative when testing against knockout controls to identify modification types. \\ 

\noindent \textbf{Detection of known and novel modifications.} Our method is built on learning from temporal signals from unmodified bases (IVT RNA), and then statistically predicting modifications that exist on native RNA signal. Biologically, this means (a) that all modifications can be measured at once on a single molecule (vs the need to generate bespoke training models per known modification that is inherent in other models) (b) there is quick adaptability to new nanopore chemistries (since only IVT data is needed for training) and (c) this approach affords the possibility not only of cataloguing known modifications on single moledules but also of discovering new modifications that are not yet understood (which can then be biologically validated).

\medskip 
We first validated our inference pipeline on well-characterized \textit{E. coli} rRNA. We then applied the method to dengue virus transcripts, discovering a novel 2'-O-methylated site in DENV subgenomic flaviviral RNA (sfRNA), orthogonally confirmed by qRT-PCR assays. We also flagged sites in longer transcripts from mammalian mRNAs after filtering out low-coverage positions in chromosome 10 and identified m6A modifications by differential anomaly rate analysis with a METTL3 knockout.

\subsection{Anomaly scores discriminate modified from unmodified RNA}\label{ssec:E_coli}

\begin{figure}[!t]
\centering
   \begin{subfigure}{0.90\textwidth}
\includegraphics[width=1.\linewidth]{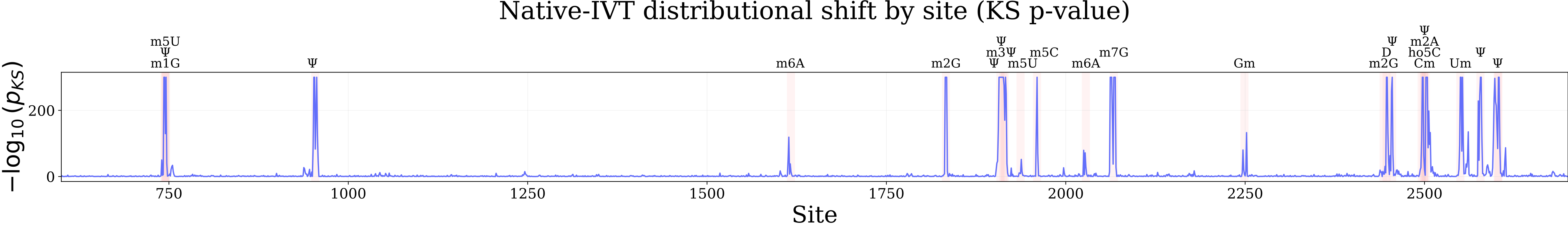}
\caption{}
\label{fig:ecoli_ks}
    \end{subfigure}
\\
   \begin{subfigure}{0.90\textwidth}
\includegraphics[width=1.\linewidth]{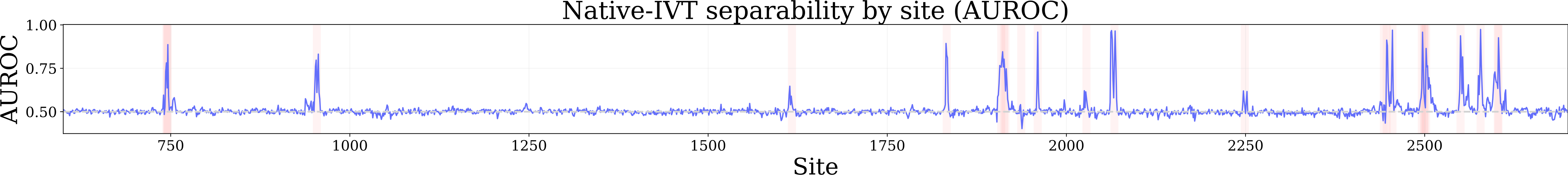}
\caption{}
\label{fig:ecoli_auroc}
    \end{subfigure}
\\
   \begin{subfigure}{0.90\textwidth}
\includegraphics[width=1.\linewidth]{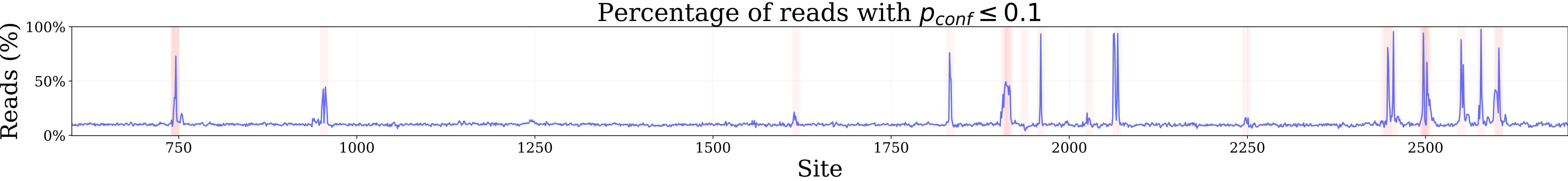}
\caption{}
\label{fig:ecoli_anomaly_rate}
    \end{subfigure}
\\
   \begin{subfigure}{0.90\textwidth}
\includegraphics[width=1.\linewidth]{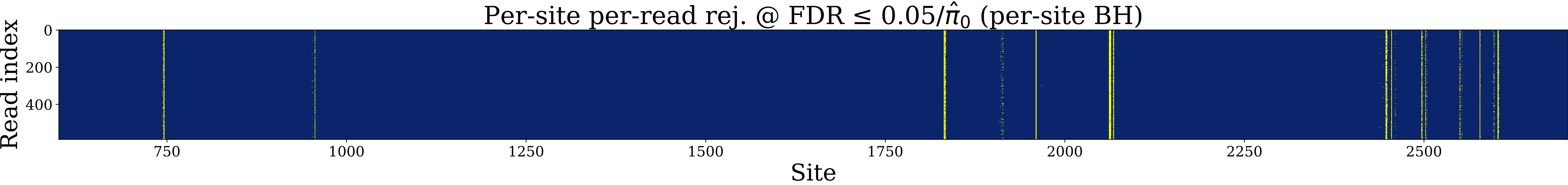}
\caption{}
\label{fig:ecoli_persite_perread_decisions}
    \end{subfigure}
\\ 
  \begin{subfigure}{0.90\textwidth}
\includegraphics[width=1.\linewidth]{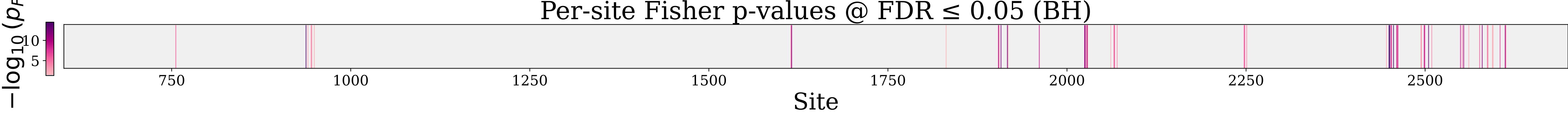}
\caption{}
\label{fig:ecoli_persite_decisions}
    \end{subfigure}
\\ 
  \begin{subfigure}{0.90\textwidth}
\includegraphics[width=1.\linewidth]{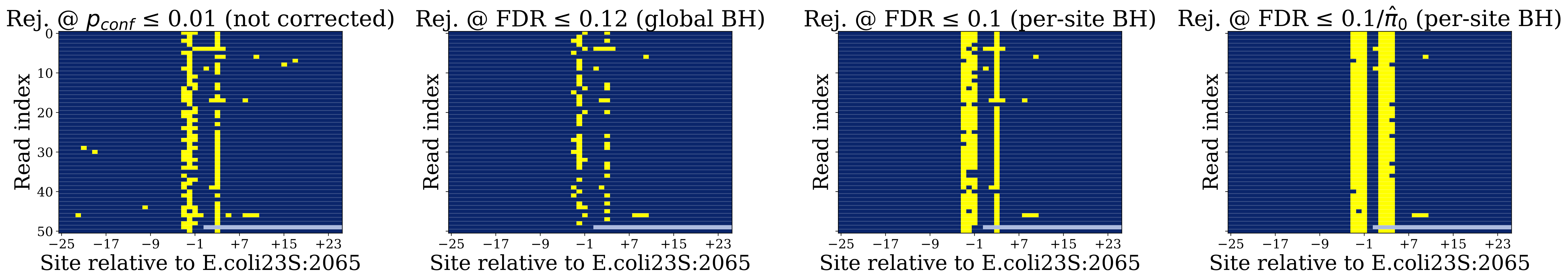}
\caption{}
\label{fig:ecoli_sub_conformal_corr}
\end{subfigure}

    \caption{\textbf{Evaluation on ribosomal RNA modifications in E. coli 23S.} (a) KS test p-values comparing the native and \acrshort{ivt} score distributions at each site. (b) AUROC values quantifying the performance of the anomaly detector. (c) Percentage of reads with
a score exceeding the 0.90 quantile of the calibration scores. (d) Single-molecule (conformal) p-values with FDR control. At a site, the conformal p-values are thresholded at the BH cutoff, further corrected with Storey's estimate $\hat{\pi}_0$ of the proportion of non-anomalous reads (e) Per-site Fisher's combination test with FDR control at level $0.05$ with \gls{bh}. The heatmap shows the Fisher p-values (light gray values are non-significant). (f) Anomaly maps obtained under different multiple testing corrections (yellow dots indicate discoveries). From left to right: using conformal p-values for each read–site pair thresholded at $0.01$; conformal p-values thresholded at the BH-adjusted level; per-site BH correction applied to conformal p-values; Storey’s procedure applied.}  
\end{figure}

We first re-analyzed the nanopore direct sequencing data for \emph{Escherichia coli} (\emph{E. coli}) ribosomal RNA (rRNA) \cite{stephenson2022direct}. These rRNAs provide an ideal benchmark with well-characterized modifications: 36 modified sites in 16S and 23S rRNA molecules exhibit 17 distinct chemical structures with known locations, types, writer enzymes, and stoichiometries \cite{siibak2010subribosomal,fleming2023direct}. The dataset (see Data availability) provides raw read FAST5 files for both \acrshort{ivt} control and native, cell-derived RNAs, both sequenced with the RNA002 chemistry. Previous computational approaches have been applied to this dataset. Nanodoc2\cite{ueda2023rna}, a deep one-class classifier trained on IVT data, learns feature representations of nanopore time series and clusters native versus IVT vectors at each site, marking sites as modified when cluster frequencies differ significantly. Another approach\cite{stephenson2022direct} applies per-site two-sample KS tests to the median current and the dwell time separately. In contrast, our approach extracts signal features via the signature transform, which captures comprehensive temporal dynamics beyond simple summary statistics, without requiring neural network training.

We computed \acrlong{nns}s using $3{,}000$ \acrshort{ivt} reads for the canonical reference data, then evaluated the discriminative power of the scores by comparing scores between a set of fresh IVT reads and a set of native reads. \Cref{fig:ecoli_ks,fig:ecoli_auroc} show the results across all 23S positions (2.9 kb), with clear peaks at known modification sites (16S: 1.5 kb results in Extended Data). For visualization, we highlight the modified five-mer (\texttt{NNANN}, where the central base is modified) and its five flanking sites in red. 

For each site, we performed a two-sided two-sample KS test comparing \acrshort{ivt} and native \glspl{nns}. The resulting $-\log_{10}(p\text{-value})$ peaks coincide with the modification sites ($\pm 5$ nt), confirming that these modifications induce distributional shifts in the \gls{nns}. As the KS test captures the score's largest distributional discrepancy, to better gauge the trade-off between sensitivity and specificity across different classification thresholds, we calculated the area under the receiver operating characteristic curve (AUROC) values, a standard metric for anomaly detection. Ideally, with \emph{read level} ground truth, the AUROC would quantify how well the \gls{nns} separates “modified” from “unmodified” reads. Lacking this ground truth, we instead treat \acrshort{ivt} versus native scores as our two classes. In this context, the AUROC is the probability that a randomly chosen native score exceeds a randomly chosen IVT score. The resulting AUROC profile shows clear separation (near $1$) nearby the known modification sites (reflecting the high modification rate) while remaining close to $0.5$ at unmodified sites, where \acrshort{ivt} and native distributions are expected to be identical. This threshold-independent metric confirms that our approach is able to rank native signals above calibration signals at modified sites.

\subsection{From anomaly scores to anomaly detection at the read level}

The aforementioned related work\cite{ueda2023rna,stephenson2022direct} operate at the \emph{site level} and neither approach provides probabilistic predictions for individual reads. While we also employed KS tests for validation purposes, our overall approach fundamentally differs by enabling genuine per-read predictions through a principled statistical pipeline. Having demonstrated that our signature-based \acrlong{nns}s effectively discriminate modified from unmodified signals across diverse rRNA modification types, next we assess the results obtained by first converting anomaly scores to calibrated p-values and then applying multiple testing corrections to achieve controlled error rates at both read and site levels.

Before this, we note that the \emph{E. coli} analysis confirmed that it is important to distinguish \emph{anomalies} from \emph{modifications}, as it is generally the case with comparative approaches. Large \glspl{nns} reflect departures from the expected (unmodified) signal distribution, arising either from genuine chemical modifications or technical artifacts such as alignment errors (Methods). Importantly, a read may still exhibit an exceptionally high \gls{nns} (we will make precise what we mean by this) at an unmodified site and be deemed anomalous. This may happen if an adjacent modification perturbs its $k$-mer signal. While a single chemical modification may only induce a modest shift in the \gls{nns} (possibly across several $k$-mers), clusters of modifications may amplify shifts. Here, we call a read ``anomalous'' at a given position if its \gls{nns} exceeds the $\alpha$-quantile (where $\alpha$ is a number close to $1$) of the empirical distribution derived from calibration (unmodified) samples. The \emph{anomaly rate} at each site is the fraction of reads flagged in this way. In \emph{E. coli}, high modification rates are expected, typically with a frequency of over 85\% when grown at 37°C \cite{fleming2023direct}—so one might expect relatively high anomaly rates at every modified site. However, because a single chemical modification can induce only a modest shift distributed across several adjacent signals—and clusters of modifications can also amplify those shifts—our anomaly rates should not be regarded as modification rate estimates but rather as \emph{detectability rates}. We found anomaly rates less than $20\%$ in the neighboring sites of m6A (1618, 2030), m5U (1939) and Gm (2251), which do not have any other modification close by (\Cref{fig:ecoli_anomaly_rate}). We found better detectability on 16S (\Cref{fig:ecoli16s} in Extended Data).

\begin{figure}[!t]
\centering
\begin{subfigure}{1\textwidth}
\centering
\includegraphics[width=0.9\linewidth]{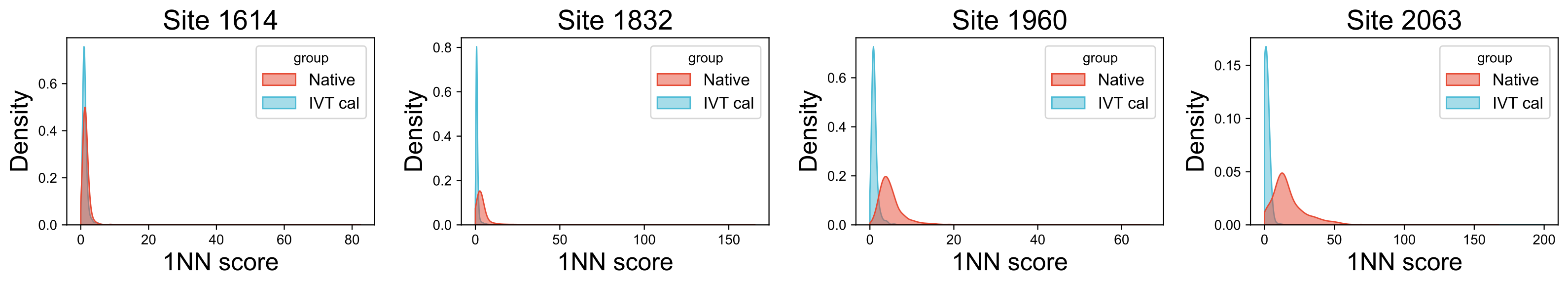}
\caption{}
\label{fig:density_plot}
\end{subfigure}
\begin{subfigure}{0.90\textwidth}
\centering
\includegraphics[width=1\linewidth]{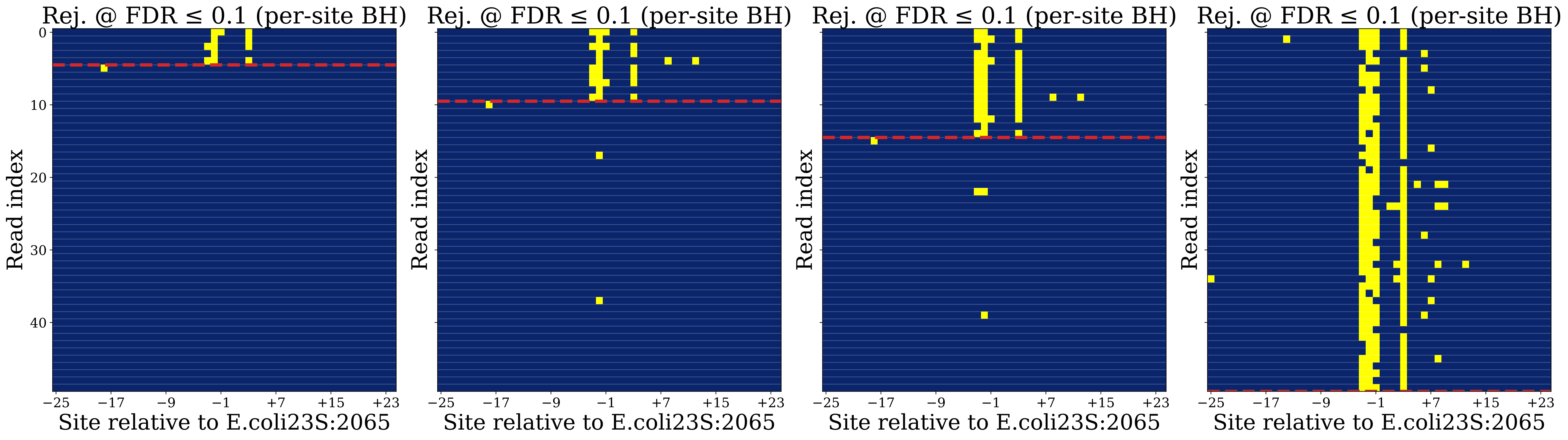}
\caption{}
\label{fig:site_level_multiple_testing_level}
\end{subfigure}
\caption{\textbf{Detecting modified sites with low stoichiometry.} (a) Density plots of IVT and native scores at four modified sites, showing the strongest shift at the last site. (b) Thresholded conformal p-values around the site harboring \texttt{m7G} for different values of \texttt{n\_cal} and modification level $x$. From left to right: $x=0.1, 0.2, 0.3, 1$. The red dotted line separates native from \acrshort{ivt} reads.}
\end{figure}

\subsection{Read-level modification detectability with conformal p-values} 

To detect anomalies in \emph{E. coli} rRNAs, we converted each \gls{nns} into a conformal p-value \cite{bates2023testing}, and flagged p-values below a prespecified threshold (first panel of \Cref{fig:ecoli_sub_conformal_corr}). Even a small regions—say, $50$ nucleotides covered by $50$ reads—entails $50\times 50=2{,}500$ tests. We therefore controlled the false discovery rate (FDR) using multiple testing corrections based on the \acrfull{bh} procedure (last three panels of \Cref{fig:ecoli_sub_conformal_corr} and Methods).

Since the conformal p-values are discrete, for a desired level, one must use a sufficiently large IVT calibration set to have a non-zero rejection probability. The calibration size needed depends on the choice of multiple testing correction. Here, we used $5{,}000$ \acrshort{ivt} calibration reads. As shown in \Cref{fig:ecoli_sub_conformal_corr}, the \gls{bh} procedure at level $0.1$ applied independently at each site is more powerful than the \gls{bh} procedure applied globally on all read-site pairs at level $0.12$. The later approach required raising the FDR level from $0.10$ to $0.12$ to start seeing some rejections.

In contrast to methods such as Nanocompore that compare an experimental sample to a control unmodified sample, our approach does not rely on high \emph{native} RNA abundance or high modification stoichiometry. Because each read yields a p-value, it can operate on single reads, making it applicable across diverse RNA species. Nonetheless, greater coverage and higher modification rates improve the precision of anomaly rate estimates and the confidence of site-level calls. When a site has a high anomaly rate (as in rRNA molecules) the \gls{bh} procedure can be overly conservative. To remedy this, we applied the \gls{bh} procedure with Storey's correction (last two panels of \Cref{fig:ecoli_sub_conformal_corr}). Because modification rates in these RNAs are high, we also tested whether the method could detect low-stoichiometry modifications by mixing \acrshort{ivt} reads with native reads in different known proportions. As shown in \Cref{fig:site_level_multiple_testing_level}, it reliably identified modified reads even at low stoichiometries ($\leq 10\%$). More generally, to detect a site with a small proportion of modified reads, and providing they induce a strong signal (the needle in a haystack problem), one needs to use a sufficiently large number of IVT reads to control the false discovery rate at a sufficiently low significance level without being overly conservative.

\subsection{Detectability of modified sites combining read-level p-values} 

To obtain a site-level p-value, we aggregated the read-level p-values at each site using Fisher's combination test, after adjusting the conformal p-values to ensure the validity of the test as further described in the Methods section. We then controlled the false discovery rate across sites at $5\%$ with \gls{bh}. Applied to \textit{E. coli} rRNAs the approach proved powerful: all known modified sites except m5U (1937) were flagged and no false positive were observed (\Cref{fig:ecoli_persite_decisions}).  By design, FDR control at $5\%$ ensures that, on average across repeated draws of native and calibration data, at most $5\%$ of discoveries are expected to be false.

\subsection{Modification discovery in the RNAs of dengue virus}\label{ssec:Dengue}

Following the characterization of our method on rRNA (which is highly modified and very abundant, simplifying modification detection), we focused on \acrlong{denv}. DENV is a single-stranded RNA virus that infects up to 400 million people annually, with approximately $100$ million developing symptoms and $40{,}000$ succumbing to severe cases \cite{niaidDengue}. It is a significant global health concern. More relevant to the context here, DENV has an RNA genome that exists as a full-length genomic RNA (gRNA) of approximately $10{,}700$ nucleotides. Beyond this primary form, a crucial element in the viral life cycle is the subgenomic flaviviral RNA. This shorter, highly structured RNA, approximately $1{,}000$ nucleotides in length, accumulates due to incomplete degradation by the cellular exonuclease XRN1\cite{chapman2014rna}. 
RNA modifications play a vital role in the DENV life cycle \cite{Ruggieri2021,Mazeud2018}, and while LC-MS studies have hinted at various RNA modifications within the DENV genome \cite{nascimentoDengueClimate}, these methods inherently lack positional resolution and are susceptible to confounding by abundant host RNA contaminants like rRNAs and tRNAs, even after depletion steps \cite{HuangA2023,Baquero‐Perez2024}. To address this challenge, Wu et al. recently developed a targeted approach to purify the full-length DENV genome from host-derived contaminants \cite{Wu2025.03.17.643699}. Using this method in combination with Oxford Nanopore direct RNA sequencing, Illumina-based bisulfite sequencing, mass spectrometry, and biological validation, they reported the discovery of a methylated cytosine (m5C) at position 1218 within the DENV gRNA, which profoundly impacts gRNA stability. We successfully applied our pipeline to the DENV gRNA and accurately detected the m5C modification at its previously validated site \cite{Wu2025.03.17.643699}. Having confirmed the detection of this known modification in gRNA, we then shifted our focus to the highly structured sfRNA. To our knowledge, this study provides the first report of an RNA modification identified within the sfRNA.

\subsection{Dengue virus full-length genomic RNA (gRNA)}\label{subsec:DENV-gRNA}

\hspace{0pt}\\Coverage in the gRNA is almost as high as in the sfRNA region for the IVT dataset, ranging from about 4,000 reads in the lower regions to nearly 20,000 reads in the higher regions. In contrast, native coverage is substantially lower, with only 30 reads in the lower regions and up to about 300 reads in the higher regions.
\begin{figure}[!t]
    \centering
\begin{subfigure}{0.55\textwidth}
\includegraphics[width=1.\linewidth]{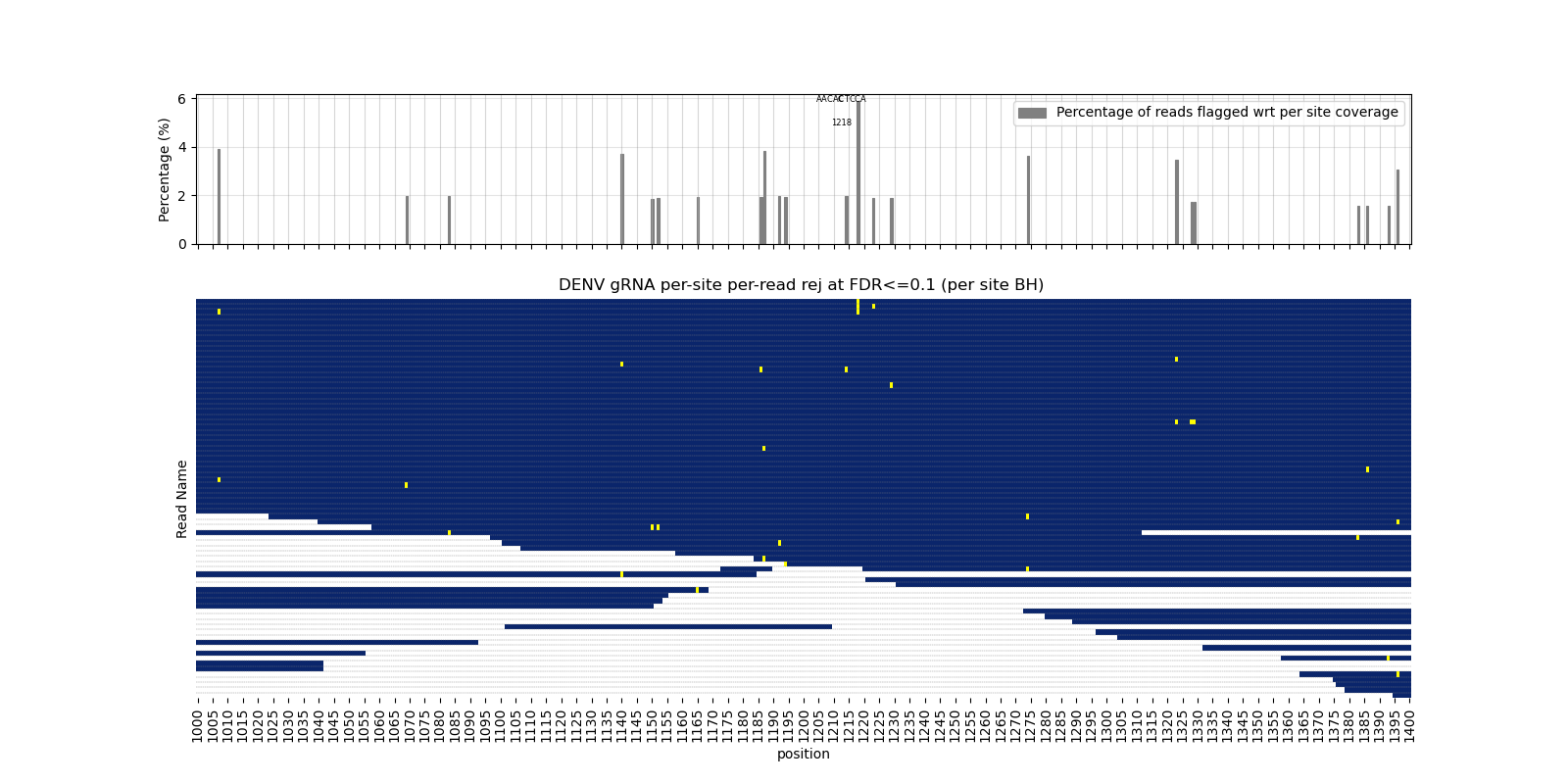}
\caption{}
\label{fig:DENVLowRegionA}
    \end{subfigure} \begin{subfigure}{0.33\textwidth}
\includegraphics[width=1.\linewidth]{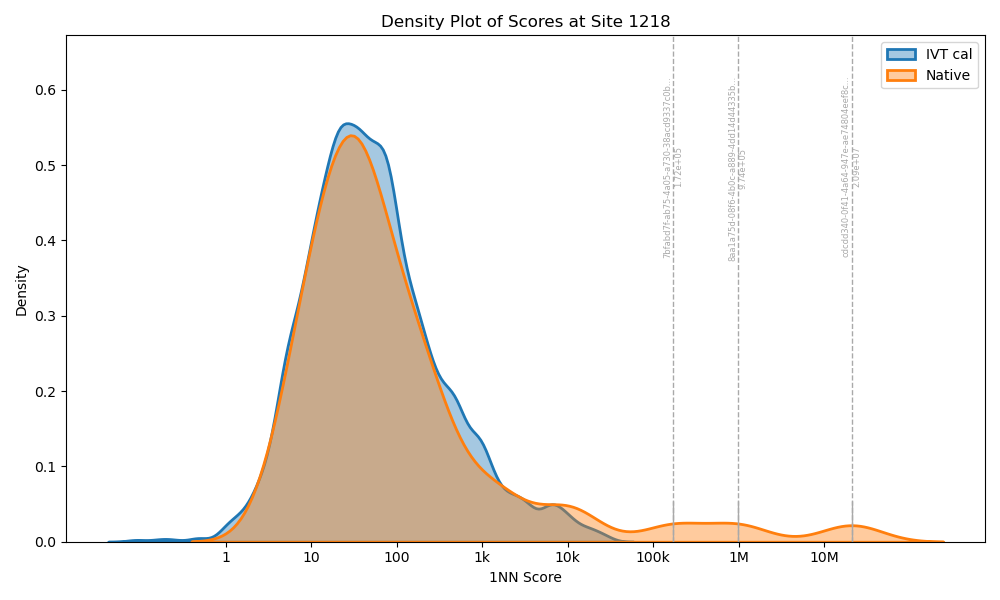 }
\caption{}
\label{fig:DENVLowRegionCDensity}
    \end{subfigure}

\vskip\baselineskip

\begin{subfigure}{0.55\textwidth}
\includegraphics[width=1.\linewidth]{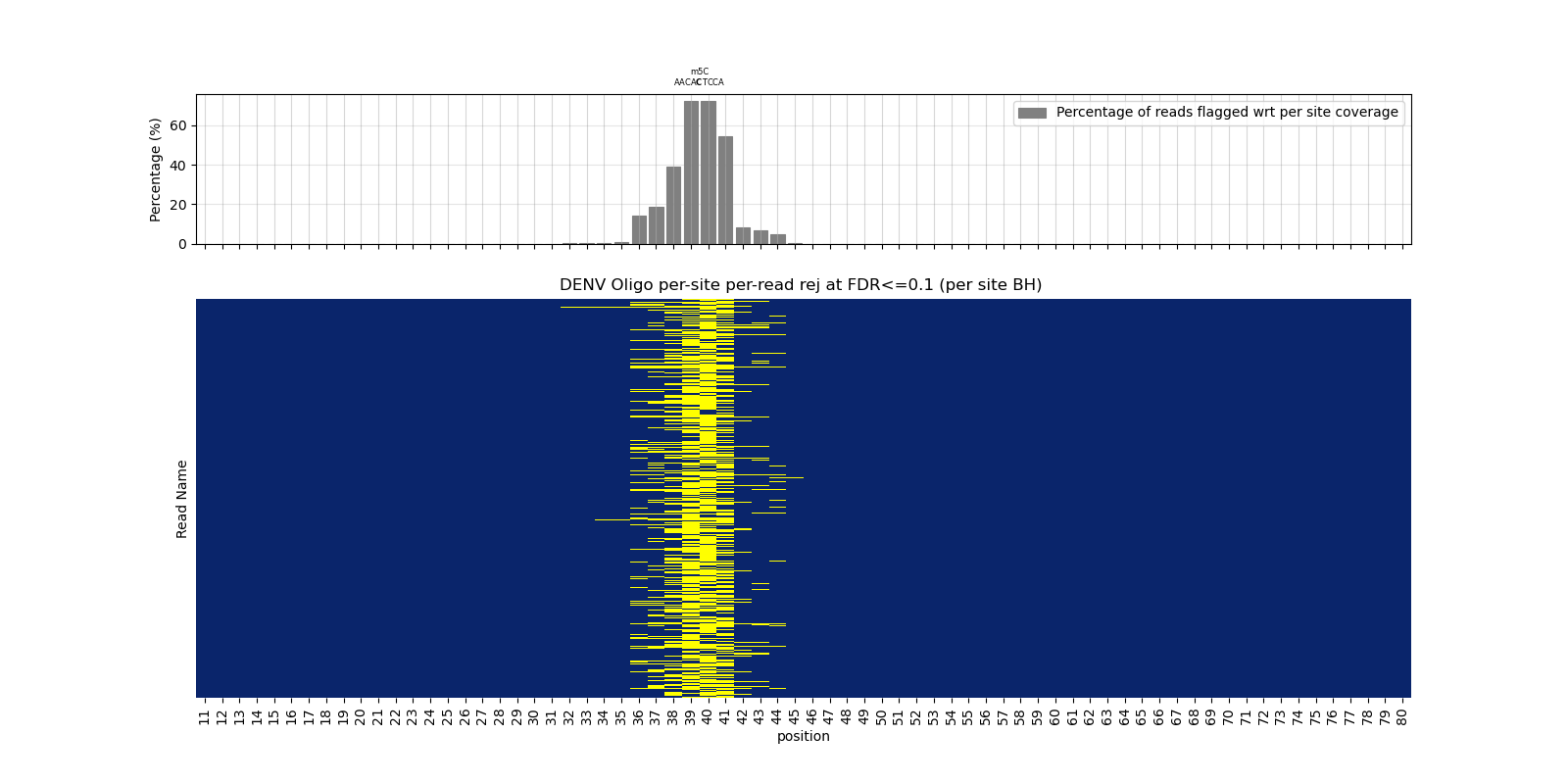}
\caption{}
\label{fig:DENVLowRegionCOligo}
    \end{subfigure}\begin{subfigure}{0.33\textwidth}
\includegraphics[width=1.\linewidth]{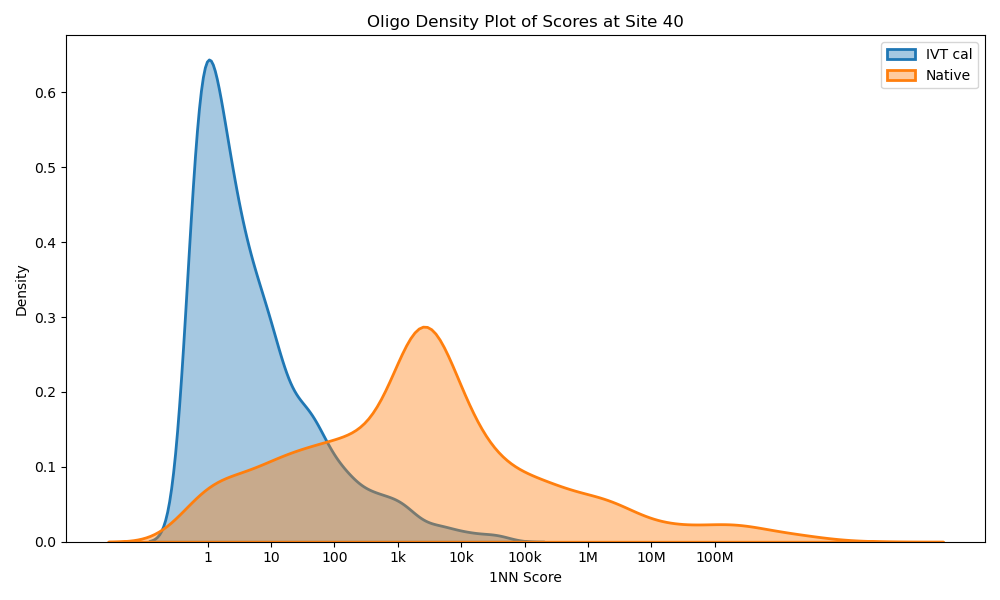}
\caption{}
\label{fig:DENVLowRegionCOligoDensity}
    \end{subfigure}

     \caption{\textbf{DENV gRNA.} (a) A 400-nt region around site 1218, previously reported as an m5C site~\cite{Wu2025.03.17.643699}. Three native reads show elevated anomaly scores at this site, with conformal p-values passing a per-site BH correction at 10\% FDR. (b) Distribution of \glspl{nns} at site 1218 for 1,600 IVT calibration reads and 50 native test reads. The two distributions are almost identical, except for the three high-scoring native reads in the tail. (c) Validation in a synthetic dengue virus (DENV) oligonucleotide bearing m5C, analyzed with the same per-site BH correction at 10\% FDR. (d) Distribution of \glspl{nns} for the oligo dataset at site 40, based on 5,000 IVT calibration reads and 2,500 native reads.}

\end{figure}

The top panel of \Cref{fig:DENVLowRegionA} shows 400 positions, with the percentage of anomalies obtained by thresholding the conformal p-values with the BH procedure at $10\%$ FDR. The bottom heatmap visualizes the underlying conformal p-values for each native read.  Our goal was to detect the m5C modification signal previously reported in this region\cite{Wu2025.03.17.643699}.  We observed an anomaly rate of approximately 6\% at position 1218, corresponding to the nine-mer sequence \texttt{AACA\textbf{C}TCCA}, with the cytosine (C) in the center being modified. \Cref{fig:DENVLowRegionCDensity} shows the density of the nearest neighbor scores for the calibration and test sets at this site. The three high-scoring reads, indicated by the grey vertical lines, correspond to the observed anomaly rate of 6\%. This m5C modification was previously reported at a frequency of about 10\% using Oxford Nanopore direct RNA sequencing data processed with the Dorado basecaller trained for m5C detection\cite{Wu2025.03.17.643699}. While their final ViREn/MiSeq bisulfite sequencing results integrate multiple experimental and computational approaches, we focus here on their Dorado-based DRS results, as these allow the most direct comparison to our method, which relies exclusively on DRS data.

Additionally, we analyzed synthetic oligonucleotide datasets containing an m5C modification at site 40 in its native sequence context, produced and provided by the authors of a prior study~\cite{Wu2025.03.17.643699}. These datasets offer a controlled experimental setting where the proportion of unmodified and modified reads can be explicitly varied, while maintaining the same underlying gRNA sequence. As expected, the distributions of the signals (and consequently their \acrshort{nns}) align with those observed in the DENV gRNA. The m5C modification was successfully detected (\Cref{fig:DENVLowRegionCOligo,fig:DENVOligoLowRegionKSAUC,fig:OligoConformalThresh}), with conformal p-values thresholded using per-site BH correction at 10\% FDR. As seen in~\Cref{fig:DENVLowRegionCOligoDensity}, the modified and unmodified reads separate clearly: the modified distribution peaks around 5,000, while the unmodified peaks near 2. In contrast, in DENV gRNA, which is only partially modified and thus not expected to reproduce the fully modified oligo distribution, we observe a small subset of high-scoring reads exceeding 100,000. This strongly supports the interpretation that these outliers represent true anomalies arising from the presence of m5C modifications.

Applying the same test across all 10,700 DENV gRNA positions, we identified two additional positions showing an anomaly rate of about 7\% (1,784 \texttt{CATCTCAAG} and 1,847 \texttt{ACAGGAAAG}), as shown in \Cref{fig:DENVgRNAIVTandTEST}. Importantly, although we had relatively few reads in the test sets, we were still able to detect a faint signal at the correct position. These additional sites are of unclear significance at this stage, but they may be interesting candidates to investigate further in future studies.

\subsection{Dengue virus subgenomic flaviviral RNA (sfRNA)}\label{subsec:DENV-sfRNA}

\begin{figure}[!t]
\centering
    \begin{subfigure}{0.9\textwidth}
        \centering
        \includegraphics[width=\linewidth]{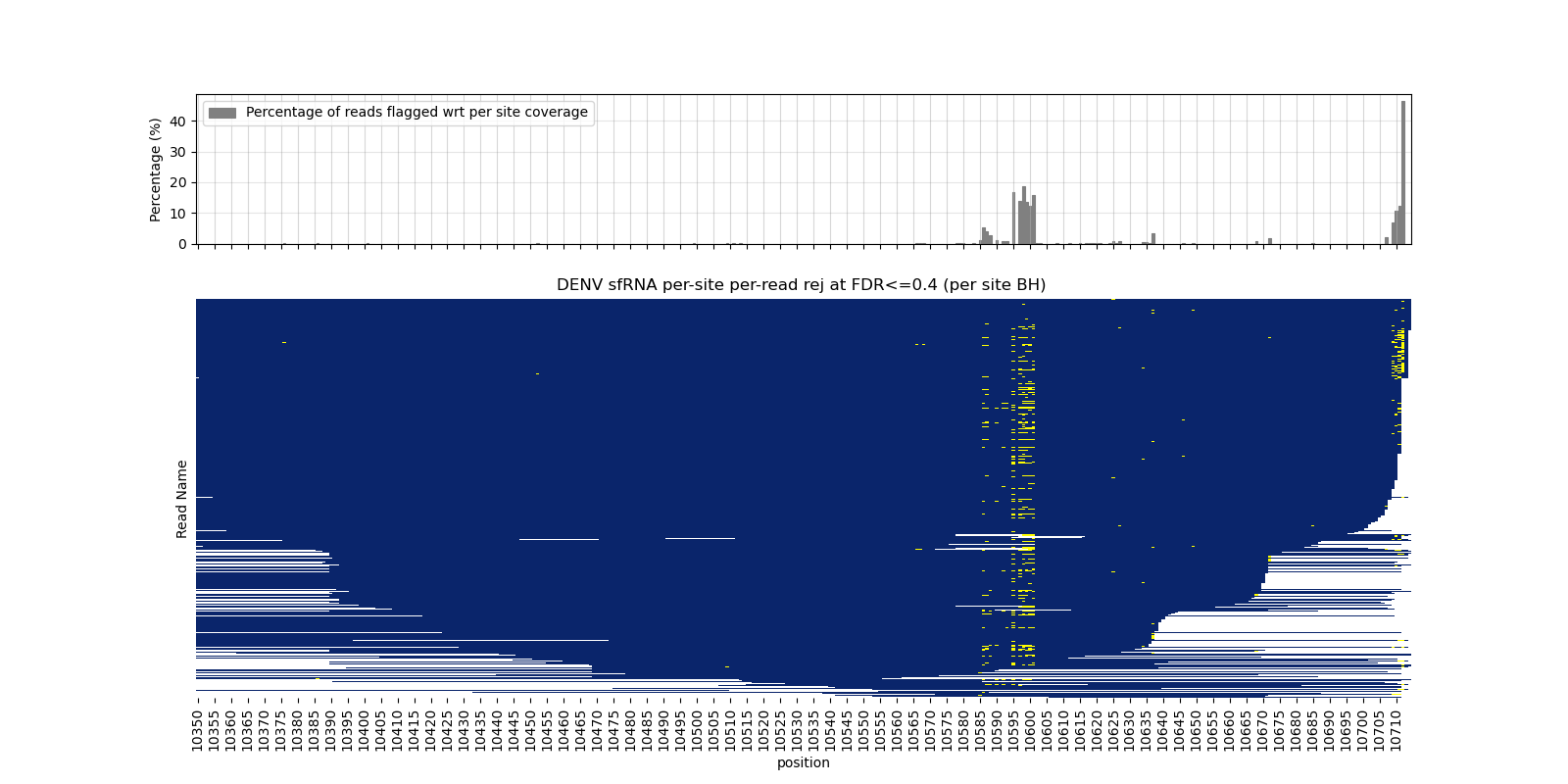}
        \caption{}
        \label{fig:BHReadssfRNAResults}
    \end{subfigure}
    \vskip\baselineskip
    \begin{subfigure}{0.68\textwidth}
        \centering
        \includegraphics[width=\linewidth]{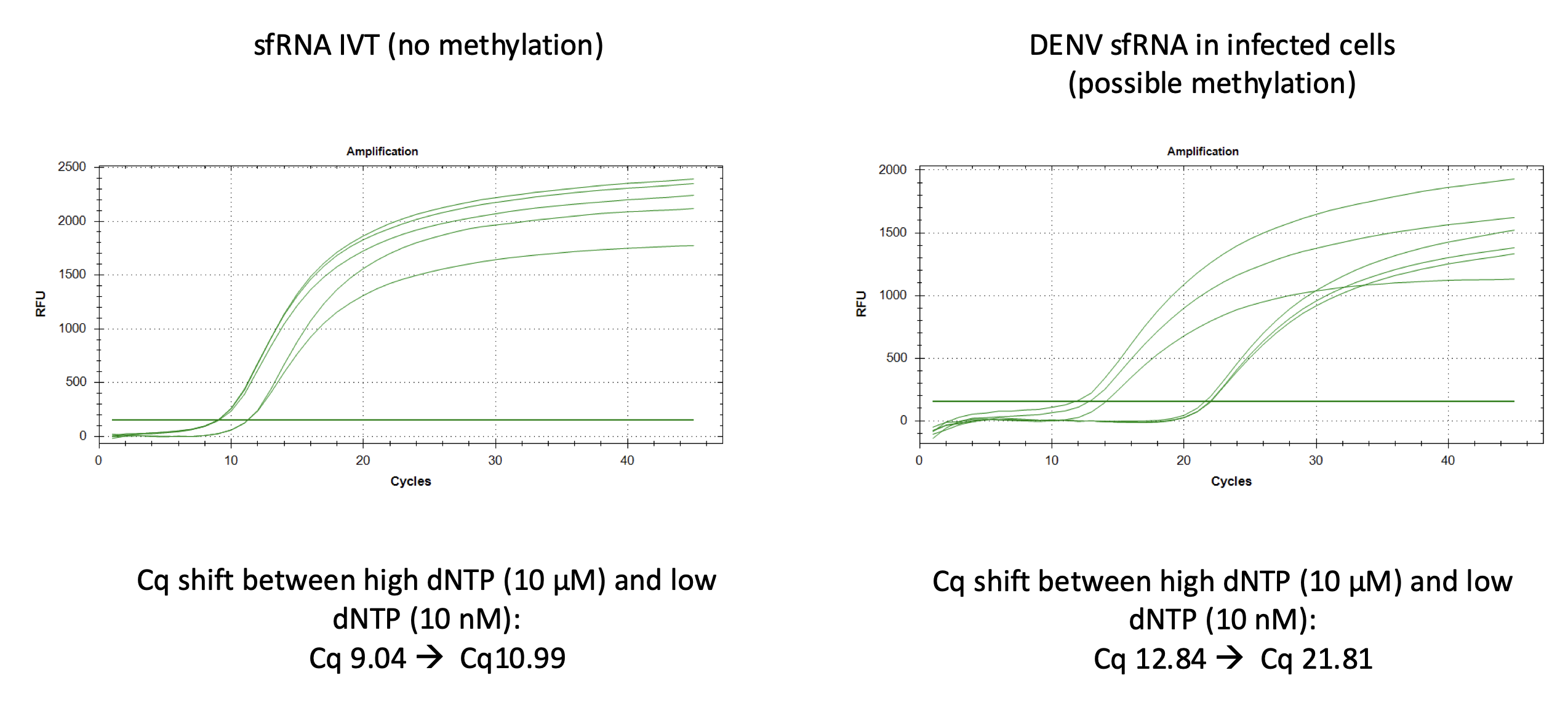}
        \caption{}
        \label{fig:DENV-Validation}
    \end{subfigure} 
    \caption{\textbf{DENV sfRNA results.} 
    Heatmap showing all reads, with conformal p-values thresholded using per-site BH correction at 40\% FDR. A clear signal can be observed around position 10,600.
    (b) qRT-PCR validation of 2'-O-methylation in DENV sfRNA. Amplification curves from qRT-PCR under low and high dNTP conditions. 
    \textit{Left:} In vitro transcribed (IVT) sfRNA shows minimal Cq shift, consistent with absence of 2'OMe. 
    \textit{Right:} sfRNA from DENV-infected cells exhibits a pronounced Cq shift under low dNTP conditions, indicative of a 2'OMe near position 10,600.
    }
    \label{fig:FullFigureDENVstructure}
\end{figure}

\hspace{0pt}\\Applying our method to the DENV sfRNA, we identified a previously unreported modification near position 10,600. 
As shown in \Cref{fig:BHReadssfRNAResults}, outlier reads cluster around position 10,600. A prior study\cite{nascimentoDengueClimate} compared RNA modification profiles across flaviviruses to give an idea of what to expect. To validate this signal, we employed qRT-PCR\cite{Elliott2021}, which exploits the sensitivity of reverse transcriptase to 2'OMe. Under low dNTP conditions, reverse transcription at modified sites is less efficient, leading to delayed amplification. Control reactions with IVT RNA lacking modifications showed minimal changes in Quantification Cycle (Cq) values—the cycle at which fluorescence first exceeds the detection threshold—between low and high dNTP conditions, confirming expected behavior for unmodified templates. In contrast, sfRNA isolated from DENV-infected cells exhibited a pronounced rightward shift in the amplification curve under low dNTP conditions (\Cref{fig:DENV-Validation}). This delay suggests the presence of a 2'OMe modification at the targeted nucleotide, as determined by the primer design used for amplification, thereby supporting our method’s predictions.

The anomalies in sfRNA impact the Dumbbell 2 (DB2) and the 3' Stem-Loop (3'SL) regions 
commonly found in the 3' UTR of flaviviruses as seen in prior studies~\cite{DENV_UTR3_0}\cite{DENV_UTR3_1}\cite{DENV_UTR3_2}, 
which is crucial for regulating various aspects of viral RNA \cite{Front_Genet_2018}\cite{whoDengueDiagnosisTreatment}. This region also serves as a site of interaction for several proteins that directly engage with RNA modification readers~\cite{Front_Genet_2018}~\cite{whoDengueDiagnosisTreatment}. The cluster of anomalies occurs in the region 10,590 - 10,600 (with reference base sequence \texttt{AGAGGAGACCCCCCCAAAAC(A)AAA}) where the 2'OMe site was found. As illustrated in \Cref{fig:Distance tree - Blast H m region}, this sequence is highly conserved among dengue virus strains and related viruses, suggesting a potential functional importance. This conservation supports the hypothesis that the observed modification contributes to a critical structural element---one that is likely essential for viral viability, as mutations in this region may prevent the virus from surviving. However, their significance is unclear and warrants further investigation.

\subsection{Modification detection in mouse messenger RNA}\label{ssec:mouse_mRNA}

\begin{figure}[!t]
\centering
   \begin{subfigure}{0.45\textwidth}
\includegraphics[width=1.\linewidth]{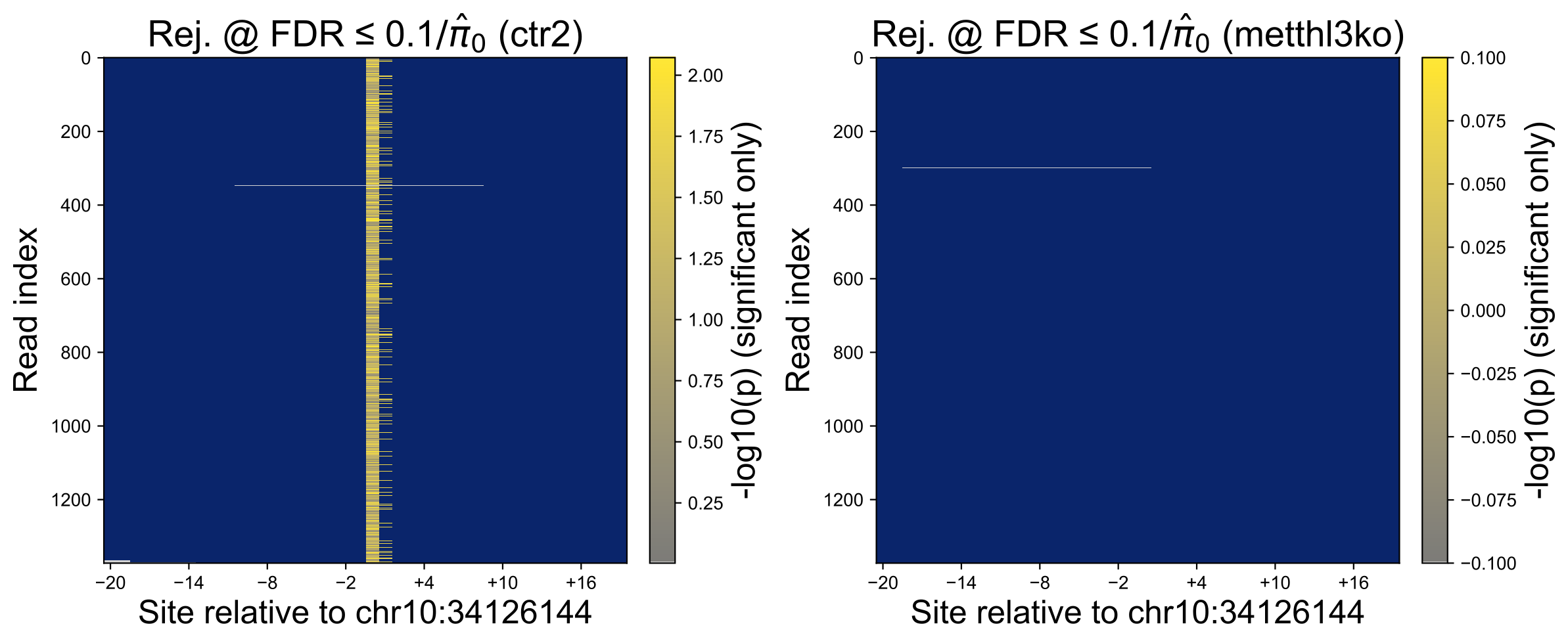}
\caption{}
\label{fig:Mouse_m6A}
    \end{subfigure}
   \begin{subfigure}{0.45\textwidth}
\includegraphics[width=1.\linewidth]{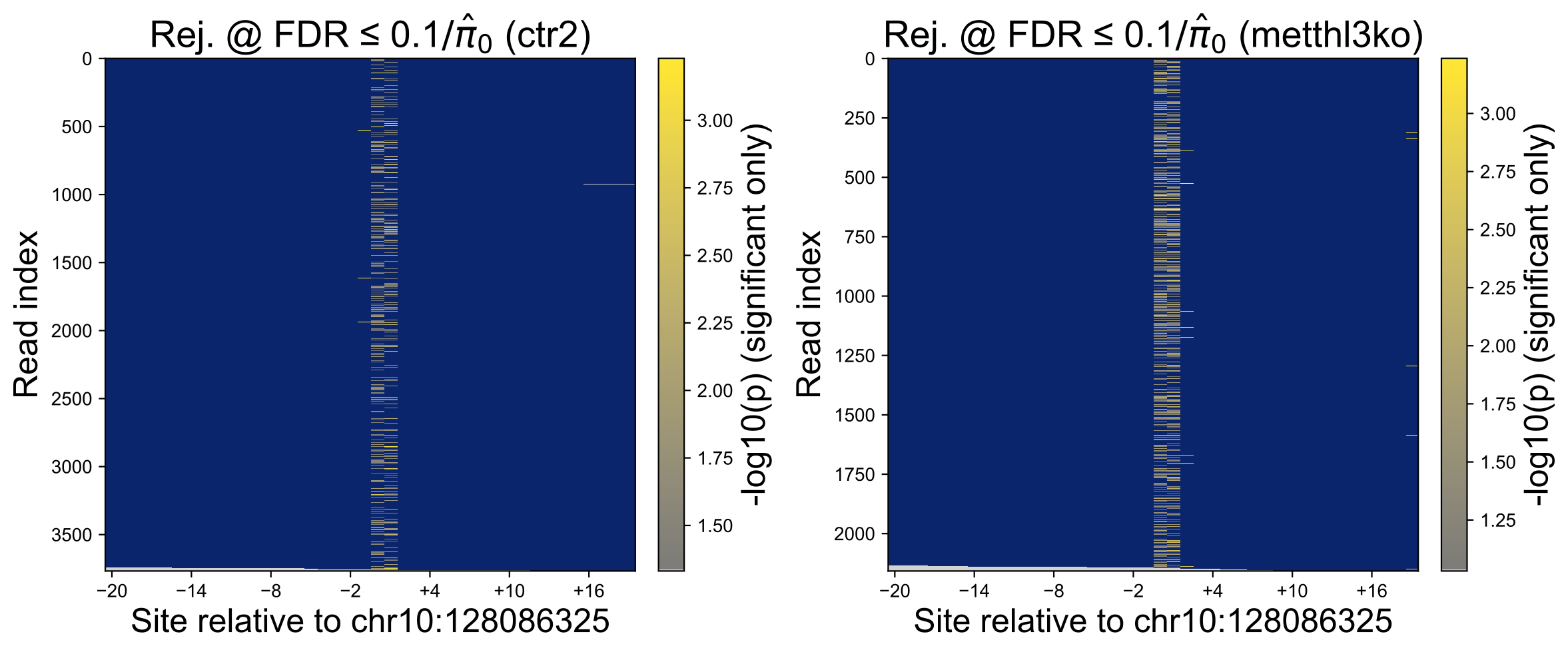}
\caption{}
\label{fig:Mouse_unidentified}
    \end{subfigure}
    \\
   \begin{subfigure}{0.9\textwidth}
\includegraphics[width=1.\linewidth]{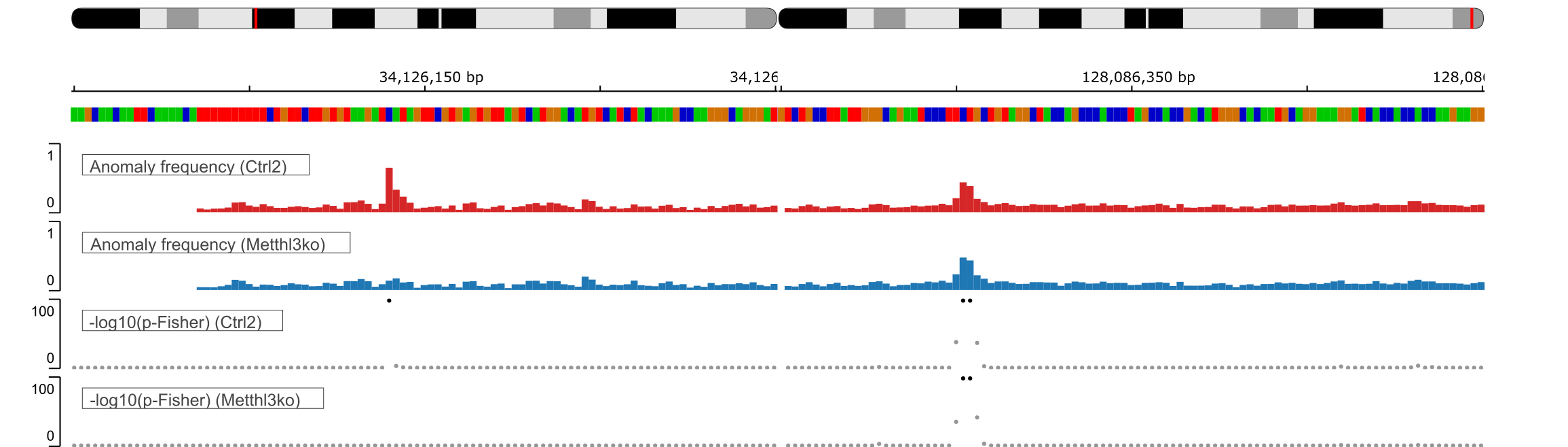}
\caption{}
\label{fig:Mouse_IGV}
    \end{subfigure}
  \begin{subfigure}{0.27\textwidth}
 \includegraphics[width=1.\linewidth]{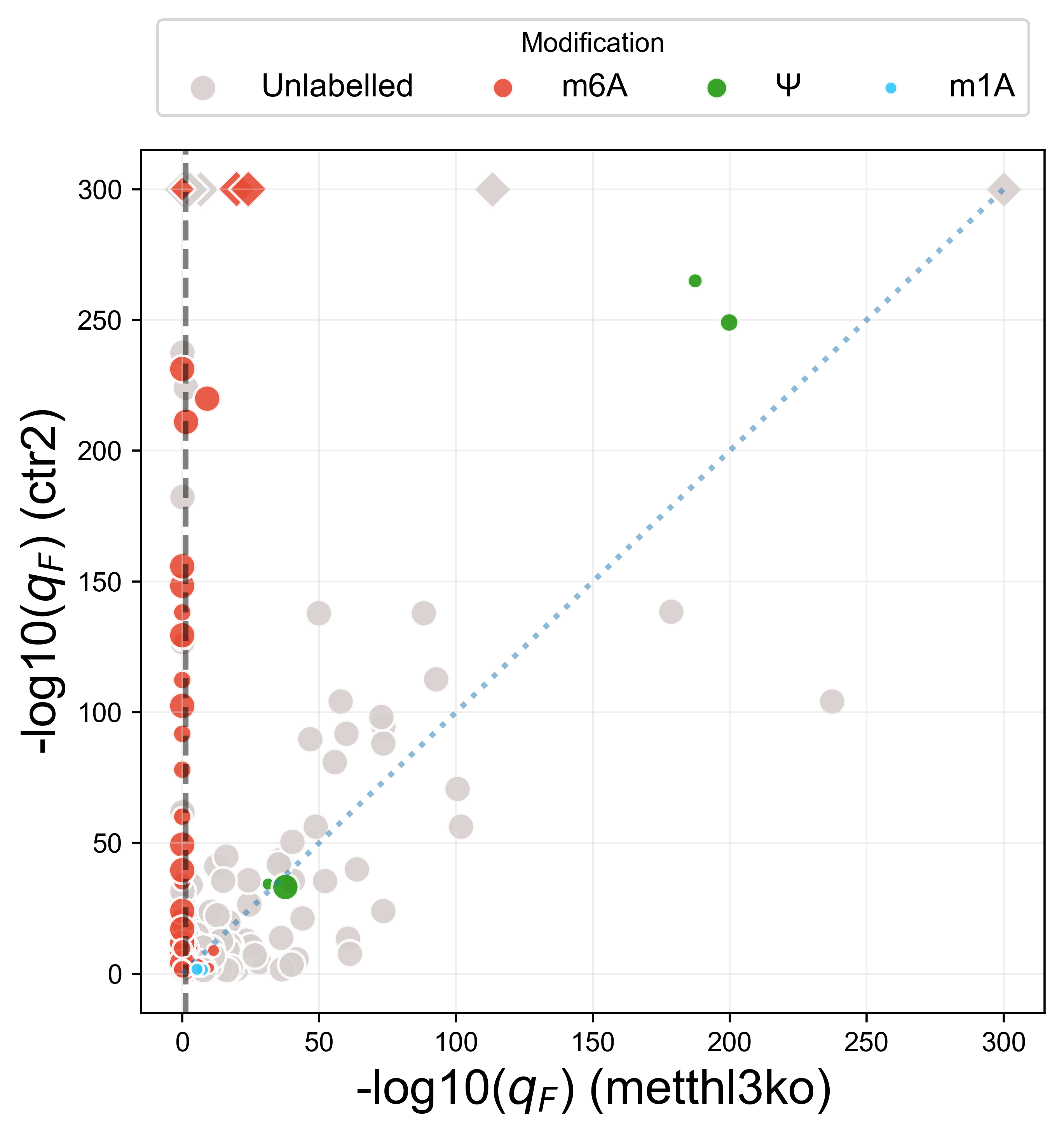}
 \caption{}
\label{fig:Mouse_Fisher_signif}
 \end{subfigure}
 \begin{subfigure}{0.27\textwidth}
 \includegraphics[width=1.\linewidth]{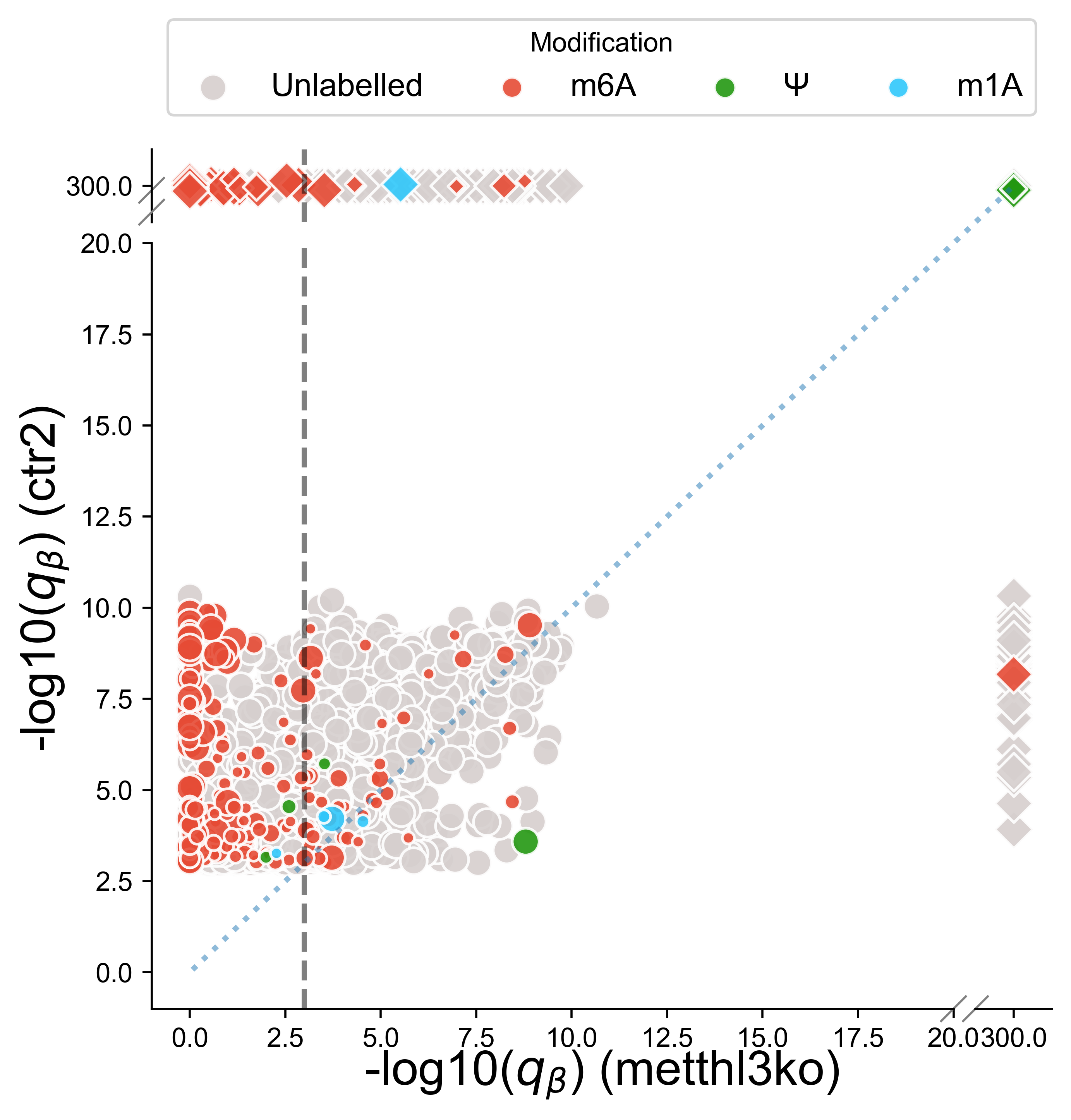}
 \caption{}
\label{fig:Mouse_BetaBinom_signif}
 \end{subfigure}
  \begin{subfigure}{0.25\textwidth}
 \includegraphics[width=1.\linewidth]{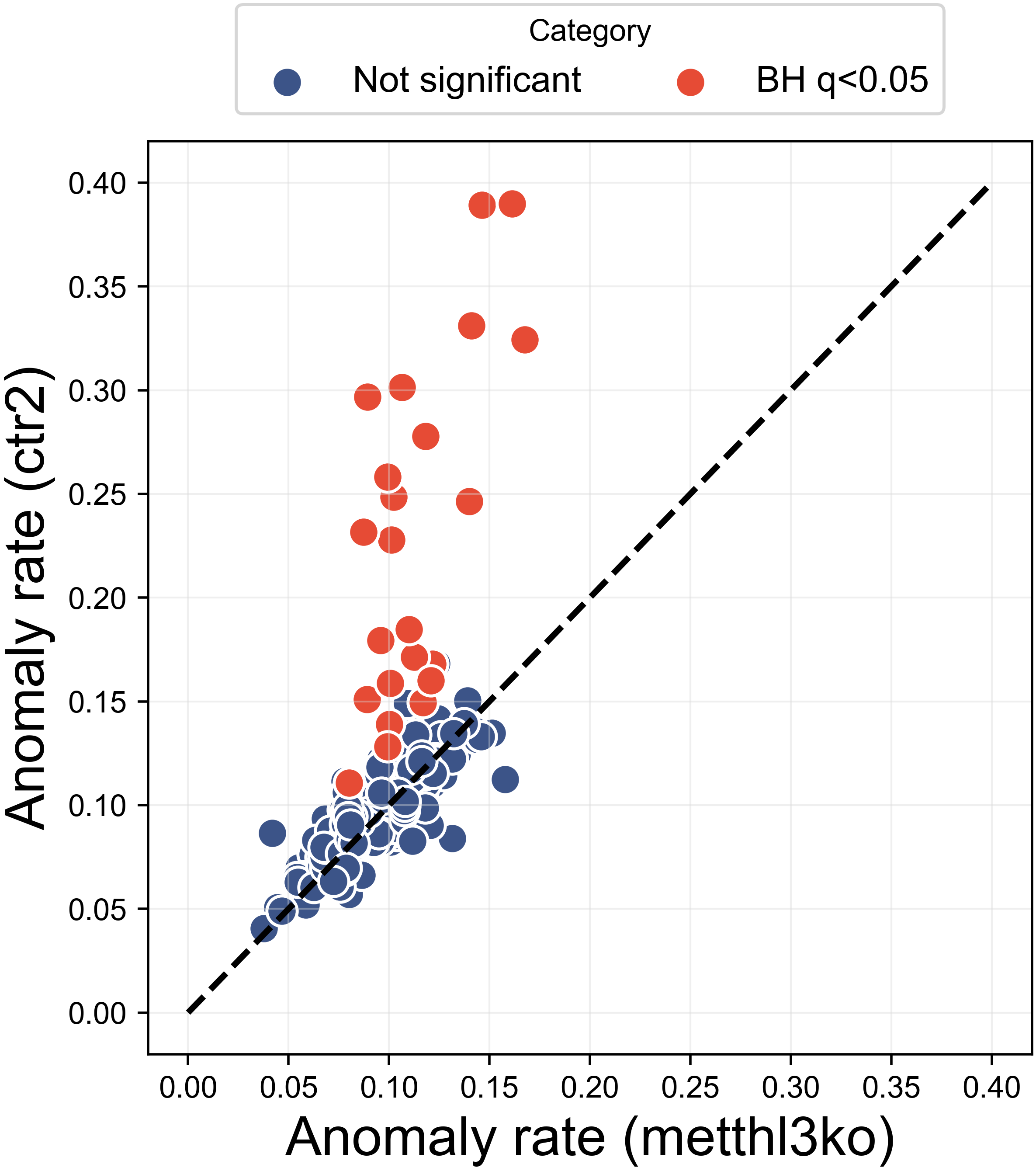}
 \caption{}
\label{fig:Mouse_DAR}
 \end{subfigure}
    \caption{\textbf{Mouse mRNA (chromosome 10) results.} Comparison of detected anomalies per read and per site between the \textit{ctr2} modified sample and the \textit{metthl3ko} enzymatic knockout. (a) Example locus where anomalies are detected in \textit{ctr2} but absent in \textit{metthl3ko} (b) Example locus where anomalies detected in \textit{ctr2} are also present in \textit{metthl3ko}. (c) \acrshort{igv} visualization of the same region on chromosome~10, showing the corresponding signal track (bedGraph) and highlighting the local enrichment of anomalies.  (d) Comparison of the effect sizes (of our detections in \emph{ctr2} using Fisher p-values corrected at level 0.05) between the modified sample and the \emph{metthl3ko} knockout. Dotted lines represent the thresholds at 0.05. Values are clipped at $10^{-300}$ (depicted by diamonds). The size of the dots represents its distance (between 0 and 4 nt) to an RMBase site (e) Same as (d) using the beta-binomial global p-value at level $0.001$. (f) Beta–binomial parametric bootstrap test for two proportions for the RMBase m6A-centered nine-mers. The significant positions are colored in red.}
\end{figure}

After our success in both detecting a very recently published modification in DENV gRNA but also, proposing and then validating a novel modification in sfRNA, we turned our attention to mRNA. This is a far more difficult task since it involves thousands of RNA species versus two for DENV (gRNA and sfRNA), and probably highly heterogenous modifications per RNA species in the population. Statistically, this problem is entirely within the remit of our approach as long as the IVT sequencing depth is adequate. To ensure robust estimation of the normal nine-mer signal distribution underlying our \glspl{nns}, we ran our tool on sites with at least $1{,}000$ IVT reads, and exploited up to $3{,}000$ reads when available. Additional IVT depth directly improves the stability of conformal p-values, reducing their variability as sample size increases. While our target depth for this was $1{,}000$ reads per site, we considered sites down to $100$ reads when necessary in the following analysis of the results. Our tool identified hundreds of anomalous sites, including a subset of robust m6A modifications that vanish in the METTL3 knockout, as well as additional sites consistent with other RNA modifications reported in RMBase.

We applied our tool to the mouse RAW 264.7 sample (Ctrl2-24h-IFN-LPS) sequenced with the SQK-RNA004 kit\cite{Val_biorxivpreprint} using the corresponding \acrshort{ivt} sample as negative control. We focused on $92{,}340$ sites on chromosome $10$ with at least one read in the modified sample (\emph{ctr2}) and covered by at least $1{,}000$ reads in the IVT control (\emph{ctr1}). Using the same \acrshort{ivt} sample and exact same pipeline, we also analyzed a METTL3 knockout sample, and therefore produced a BED file for two samples (\emph{ctr2} and \emph{metthl3ko}). For the following analysis, we only considered sites with more than $10$ reads and nine-mers that do not contain single-nucleotide variants, yielding $86{,}787$ analyzable sites in \emph{ctr2}. 

As an initial exploratory step, we examined the sites flagged as anomalous by our tool. To this aim we further restricted the \emph{ctr2} BED to sites covered by at least $100$ IVT-calibration reads, yielding $30{,}680$ sites, each with an associated global Fisher p-value. Applying the BH procedure at a $5\%$ FDR level to these p-values we identified $237$ significant sites. We then visualized read-level conformal p-values in small genomic windows around a subset of these sites and contrasted the patterns between the modified and knockout samples. As shown in \Cref{fig:Mouse_m6A} and \Cref{fig:Mouse_unidentified}, some of the detected sites exhibit significant conformal p-values only in the \emph{ctr2} sample, suggesting that these signals are induced by an m6A modification (METTL3-dependent). Other sites show significant p-values in both samples, indicating the presence of additional modification types independent of METTL3.

As an external validation, we cross-referenced our calls with RMBase \cite{sun2016rmbase, xuan2018rmbase}. To enable comparison in our nine-mer representation, we shifted each RMBase site by 4 nt centering the putative modified adenine (A)---or thymine (T), depending on the strand. Among the $86{,}787$ analyzable sites, there were $845$ matches with the RMBase m6A entries. Of these, $294$ had at least $100$ calibration reads on the IVT sample and were therefore included in our Fisher test. Intersecting our $237$ discoveries with this set, we found $27$ nine-mers that were significant and located within four nucleotides of an RMBase nine-mer bearing a central m6A. For most overlapping sites, the knockout sample was not significant (see \Cref{fig:Mouse_Fisher_signif}), consistent with an m6A-dependent effect. Only two pseudouridine ($\Psi$) and five m1A from RMBase were present in the set of sites included in our Fisher test, and of these, one $\Psi$ and two m1A were significant. These modifications are not expected to be removed by the METTL3 knockout, which aligns with our observations \Cref{fig:Mouse_Fisher_signif}. We repeated the analysis with an alternative test that models the counts of anomalous reads at a site using a beta–binomial distribution with parameters determined by calibration size, test size, and threshold (Methods). As shown on \Cref{fig:Mouse_BetaBinom_signif}, this test detected more RMBase-annotated sites overall ($168$ m6A, $2$ m1A and 3 $\Psi$), but it also identified more sites that remained significant in the knockout sample. In both figures, point size encodes the absolute offset (0–4 nt) between our called nine-mer and the RMBase centered nine-mer (the closest the bigger).

Finally, we examined the \emph{differential anomaly rate}, our (modification-agnostic) analogue of the \emph{differential methylation rate} commonly used for m6A studies. At each site, anomaly counts follow a beta–binomial distribution (Methods), which supports a one-sided two-proportion test comparing \emph{ctr2} to \emph{metthl3ko}. We implemented a beta–binomial test with parametric bootstrap calibration of the null (Methods) and applied it to the $294$ RMBase sites with more than $100$ calibration reads. As shown on \Cref{fig:Mouse_DAR}, only a subset was significant ($22$ sites) after \acrshort{bh} correction at level $0.05$, indicating that elevated anomaly rates in \emph{ctr2} are not systematic across RMBase-annotated loci. This could be due to inaccuracies in RMBase annotations or a limited sensitivity of our method at certain sites, which could be improved with deeper sequencing.

\section{Discussion}\label{sec:discussion}

\medskip 

\subsection{Summary}
We introduced a computational tool to detect atypical ionic current signal stretches in high-throughput nanopore direct RNA sequencing data using the signature transform. Our tool outputs a \acrlong{nns} at every site along every read in a sample and converts these into empirical p-values to simultaneously detect multiple RNA modifications at both individual read and aggregated site levels. The resulting profiles of statistically significant deviations from modification-free measurements enable the prioritization of sites for downstream analysis by modification-specific classifiers or orthogonal biochemical validation. We validated the specificity and sensitivity of our approach across a broad range of modification types using well-characterized \textit{E. coli} rRNA molecules. Applying our method to new datasets, we identified a previously unreported 2'-O-methylated site in the 10,600 region within the highly structured Dengue virus sfRNA. During this study, the SQK-RNA004 sequencing kit became more broadly adopted by the research community\cite{zou2025comparative}, replacing the SQK-RNA002 kits. Our framework proved robust across both kits, highlighting a key advantage of our parameter-free approach: no retraining is required with \acrshort{ont} software or chemistry updates.

\subsection{Key features}

Our tool, which operates as an anomaly detector rather than a supervised classifier trained on specific modifications, complements existing modification detection pipelines.

It can be applied to any RNA sample of interest, requiring only modification-free reference data produced in sufficient quantities by \acrlong{ivt}. 
Specifically, the larger the number of \acrshort{ivt} reads, the better our method can distinguish truly unmodified sites (\glspl{nns} cluster near zero) from potentially modified ones. Additionally, a large number of (calibration) \acrshort{ivt} reads increases the resolution of empirical p-values, reducing the variability of the false positive rate across different experiments hence providing more reliable statistical confidence. The single limitation of our method is currently the depth of IVT sequencing data we have in hand (the deeper the IVT data the better the ability to predict modifications), which is easily surmountable especially if there is concerted effort by the community toward this.

As a parameter-free method, our framework requires no adjustments when sequencing chemistry changes. While newer chemistries may produce different modification signals, our approach automatically adapts by comparing experimental samples to chemistry-matched canonical references.

Unlike methods designed to detect specific modifications (like m6A or m5C), our approach identifies any stretch of nanopore signals that deviates from normal behavior, enabling the simultaneous detection of sites decorated by different RNA modification types. While it cannot specify which modification is present, it significantly narrows down regions requiring further analysis, thereby enabling the integration of complementary techniques, such as mass spectrometry, or modification-specific detectors. 

\subsection{Areas of improvement}

In this study, for each RNA sample analyzed, we collected \acrshort{ivt} reads aligned to the same reference sequence. Additionally, to compute the \acrlong{nns} at a given site, we used unmodified current stretches from that exact genomic position. This site-specific approach requires a large number of \acrshort{ivt} reads covering each transcript of interest, which may present some limitations. Computationally, it prevents precomputation and reuse of reference datasets across studies, requiring the recalculation of the nearest neighbor metric for each new analysis. Practically, it necessitates generating fresh \acrshort{ivt} for every sample, increasing experimental costs. Initially, we considered building universal reference datasets based on $k$-mer sequences, assuming that identical $k$-mers would produce similar nanopore signals regardless of their genomic location. However, our analyses revealed that identical $k$-mers at different positions often exhibit distinct signal distributions (see \Cref{fig:MMD_test} in Extended Data). This sequence context dependency reflects the complex biophysics of nanopore sequencing, where local sequence context affects translocation dynamics. For example, studies have previously highlighted the motor protein used in RNA002 sequencing kits has a tendency to pause on guanosine-rich sequences (i.e. the dwell time increases), attributing this behavior to high single-stranded stacking energies that might decelerate translocation \cite{stephenson2022direct}.

In the mouse mRNA analysis, we observed that with the uneven \acrshort{ivt} coverage we could analyse approximately 2\% of the sites in the experimental sample. Improving \acrshort{ivt} coverage uniformity would enable broader transcriptome-wide analysis. Several strategies could enhance coverage breadth, including pooling multiple \acrshort{ivt} preparations, or optimizing the multi-step \acrshort{ivt} protocol to reduce directional coverage bias along transcript length. 

We used the signal-to-reference alignment mode in Uncalled4.
While this alignment mode may introduce errors due to genetic variations between samples and the reference sequence, it offers a shared coordinate system (to align the experimental sample and the \acrshort{ivt} samples) and independence from basecalling errors that could confound our anomaly score calculations. Signal-to-read alignment represents an intriguing alternative that could mitigate reference-induced alignment errors. As signal alignment algorithms continue to improve, we anticipate corresponding improvements in our anomaly detection sensitivity and specificity.

\section{Methods}

\subsection{Nanopore signal alignment}
Nanopore signal alignment (also known as event alignment, segmentation, or resquiggling) provides the necessary starting point for our subsequent analyses. It allows us to segment the raw nanopore signal into shorter stretches each in correspondence with a $k$-mer in the reference nucleotide sequence, and select the appropriate unmodified reference data to compute nearest neighbor scores. It also allows us to map the anomalies detected by our tool to a precise genomic or transcriptomic location. 

 We use the software \texttt{Uncalled4} \cite{kovaka2024uncalled4} to do this signal alignment. It first translates the reference sequence into an expected current profile using a \emph{$k$-mer pore model}, and then aligns the read signal to this reference profile using a dynamic programming algorithm. More formally, the reference sequence is a polymer $$\poly{b}=(b_1,\ldots,b_N),$$ where each unit $b_i$ is A,C,G,T (or U) in the case of canonical (unmodified) RNA. A subsequence $\poly{b}_{m:n}=(b_m,\ldots, b_n)$ of length $n-m+1=k$ is a \emph{$k$-mer}. Let $0=t_1<t_2<\ldots<t_\ell=T$ and denote by $$X=(X_{t_1}, \ldots, X_{t_\ell})$$ the sequence of real-valued ionic current measurements $X_{t_i}\in \mathbb R$ as a polymer passes through the pore over the time period $[0,T]$. Nanopore signal alignment partitions the time indices $\mathcal{T}=\{t_1, \ldots, t_\ell\}$ into disjoint sets \(\mathcal{T}_1, \ldots, \mathcal{T}_N\) such that $\bigcup_{i=1}^{N}\mathcal{T}_i=\mathcal{T}$. Each segment $$X^{(i)}:=\{X_t : t \in \mathcal{T}_i\}$$ corresponds to a particular $k$-mer at position $i$ in the reference sequence $b$. This segmentation provides the basis for extracting localized features from the signal trace.

Alignment tools rely on so-called \emph{$k$-mer pore models}. In nanopore sequencing, a motor protein controls the molecule's passage through the nanopore by advancing it in discrete, stepwise increments. The reader head, being larger than a single nucleotide, measures an ionic current generated by the ensemble of bases (the $k$-mer) present in the pore at each step. Each $k$-mer can be associated with summary statistics (typically the mean and standard deviation) of the current features (such as mean, standard deviation and dwell time), which defines a pore model. These $k$-mer statistics, computed over multiple sequence contexts, depend on the sequencing chemistry and ONT software\cite{kovaka2024uncalled4}.\\

\subsection{Nanopore direct RNA data preprocessing} 
Sequencing of Dengue sfRNA was performed on a PromethIon platform utilizing flowcell FLO-PRO004RA and SQK-RNA004 kit.
Dorado V0.8.1 (GPU mode) and minimap2 were used for basecalling and alignment, followed by indexing and sorting of the aligned BAM files with samtools.\
Basecalling was performed with following Dorado parameteres: 

\begin{verbatim}
dorado basecaller \
--min-qscore 6 \
--emit-moves \
sup /pod5_path/ \ 
--reference /Reference.fa > /File.bam
\end{verbatim}

The resulting BAM file was then sorted and indexed using samtools:

\begin{verbatim}
samtools sort \
-@ 15 \
-o File.sorted.bam /File.bam &
samtools index \
-@ 15 /File.sorted.bam
\end{verbatim}

Signal alignment was performed using Uncalled4 v4.1 as follow:

\begin{verbatim}
uncalled4 align \
--ref /Reference.fa \
--reads /pod5/ \
--bam-in /File.sorted.bam \
-p 10 \
--eventalign-out \
--eventalign-flags samples,print-read-names \
--ordered-out | \ 
gzip > File.txt.gz
\end{verbatim}

\medskip 
\subsection{Data filtering}
A \emph{basecalled read} is a contiguous sequence of nucleotides (A, C, G, T or U) produced by basecalling an ionic current trace generated by the sequencing device as it decodes a fragment of DNA or RNA. Each read typically corresponds to a single molecule passing through the pore or being sequenced. While multiple reads may cover the same genomic region, they originate from independent molecules. Reads should generally be discarded only if their quality is poor. However, overly aggressive filtering can lead to a loss of resolution, a reduced data complexity, and ultimately an underestimation of modification frequencies. Therefore, we applied a minimal yet principled filtering strategy.  

\medskip 
\noindent\textit{Inconsistent alignment.} In the alignment file, the \texttt{reference\_kmer} field is expected to match the \texttt{model\_kmer} field. We observed instances where this was not the case. Rather than discarding the entire read, we only excluded the affected positions from further analysis.

\medskip 
\noindent\textit{Signal duplication artifacts.} We identified consecutive positions (different $k$-mers) associated with identical current values. These regions are likely artifacts or alignment errors and are excluded from the dataset. 

\medskip 

\noindent\textit{Known mutations (SNVs).} Positions flagged as mutations in the \texttt{bcftools} output. If they have an allelic fraction (AF) $\geq 5\%$ and total coverage $\geq 10$ reads, these are likely true single nucleotide variants (SNVs) rather than modification-induced artifacts. Sites where anomaly detection overlaps with single-nucleotide variants (SNVs) are flagged in the BED methyl file as SNV-positive. Downstream analyses are performed only after identifying SNVs and removing them from the dataset.
\medskip 

\subsection{Feature extraction}

Although the sequencer measures the current at a fixed sampling frequency, molecules traverse the nanopore at variable speeds. Consequently, the number of measurements per base, as determined by nanopore signal alignment, is variable. To handle this variability, previous studies either extract simple handcrafted features---such as median current and dwell time \cite{leger2021rna}---or apply interpolation and resampling techniques \cite{wu2024transfer, vujaklija2025detecting}. Here, we use the \emph{signature transform}\cite{chevyrev2016primer,perez2018signature,morrill2020generalised,lemercier2021distribution,fermanian2023new,lyons2025signature}, a feature extractor specifically tailored for sequential data and particularly effective for handling variable-length multivariate time series. We use the highly optimized Python library \texttt{iisignature} \cite{reizenstein2018iisignature} to compute signatures of nanopore ionic current time series.

\medskip 
\noindent\textit{Signature transform.} The signature $\phi(\pathx|_{[s,t]})$ of a smooth path $\pathx$ over the interval $[s,t]$ taking its values in $\mathbb R^d$, is a series of tensors in $T((\mathbb{R}^d))=\mathbb R \times \mathbb R^d\times (\mathbb R^d)^{\otimes 2}\times \ldots \times (\mathbb R^d)^{\otimes m}\times \ldots $,
\begin{equation}
    \phi(\pathx|_{[s,t]})=(1,\phi_1(\pathx|_{[s,t]}),\phi_2(\pathx|_{[s,t]}), \ldots, \phi_m(\pathx|_{[s,t]}), \ldots)
\end{equation}
where the $m^{th}$ tensor ($m>0$) is given by the following iterated integral valued in $(\mathbb R^d)^{\otimes m}$
\begin{align*}    \phi_m(\pathx|_{[s,t]})=\underset{s< u_1<\ldots<u_m<t}{\int \ldots \int} \pathx'(u_1)\otimes \ldots \otimes \pathx'(u_m)du_1\ldots du_m.
\end{align*}
Truncating it at order $m$ gives a finite feature vector of size $d'=\sum_{k=0}^{m}d^k=(d^{k+1}-1)/(d-1)$. The complexity of computing the signature of a piecewise linear path with $N$ increments (equivalently, a length-$N$ time series) is $\mathcal{O}(Nd')$.

\medskip 
\noindent\textit{Preprocessing the ionic current.} We preprocessed the ionic current (by adding a monotonically increasing coordinate, and applying the invisibility transform \cite{yang2022developing, chevyrev2016primer, morrill2020generalised, fermanian2021embedding}) to ensure that the signature transform is injective which guarantees that distinct signals produce distinct signatures. Furthermore, the universality property\cite{kidger2019deep} of the signature feature map ensures that any variation in the signal caused by a modification can be captured by a linear combination of signature terms. As we truncate the signature to obtain a finite dimensional feature vector, we potentially discard components that discriminate the signal of a modified polymer. However, the higher order terms of the signature that we omit capture finer, localized details  of the underlying path and we hypothesize that beyond a certain order, these terms are irrelevant for discriminating modifications. The results on rRNAs and mRNAs were obtained by truncating the signature at level $m=3$, and the DENV results at level $m=4$. Compared to commonly used features like dwell time and median current, the signature transform offers a more comprehensive representation of the signal dynamics. However, it is possible readout these features from the signature.

\medskip 
\noindent\textit{Dwell time.} The signature features of a one-dimensional signal are trivial, as they correspond to scaled powers of the increment $X_{t_\ell}-X_{t_1}$. To circumvent this, we augment the signal with extra coordinates. For example, we transform the nanopore time series $X=(X_{t_1}, \ldots, X_{t_\ell})$ into $\widetilde{X}=(\widetilde{X}_{t_1}, \ldots, \widetilde{X}_{t_\ell})$, where $\widetilde{X}_{t_i}=(t_i, X_{t_i})$. The first order term of the signature is the increment, which in this case, is given by $(t_\ell-t_1, X_{t_\ell}-X_{t_1})$. The first coordinate corresponds to the dwell time. 

\medskip 
\noindent\textit{Mean or median current intensity.} The invisibility transform ensures that the initial measurement is retained. The signature of the invisibility transformed path contains the terms of the original signature. The arithmetic mean $m=\frac{1}{\ell}\sum_{i=1}^{\ell}{X_{t_i}}$ can be viewed as an approximation of $\frac{1}{t_\ell-t_1}\int_{t_1}^{t_\ell}f(t)dt$, which is a second order iterated integral of $\widetilde{X}_t=(t/(t_\ell-t_1), f(t))$ when $f(0)=0$. The signature we compute corresponds to a trapezoidal approximation of this integral. When the number of samples increases, the arithmetic mean and the trapezoidal approximation (signature of piecewise linear interpolation path) converge to the integral (time average of the nanopore current). We can always get $f(0)=0$ by shifting the signal $t\mapsto f(t)-f(0)$. The signature remains unchanged, so we don't need to do this shift in practice. All in all, to retrieve the mean value with a linear combination of signature terms, one can transform the time series $
X=(X_{t_1}, \ldots, X_{t_\ell})$ with values in $\mathbb R$ into a new time series with values in $\mathbb R^3$, 

\begin{align}\label{eq:path_aug_1}
    \widetilde{X}=\left(\begin{pmatrix}X_{t_1} \\ 1/\ell\\ 1\end{pmatrix}, \begin{pmatrix}X_{t_2} \\ 2/\ell\\ 1\end{pmatrix} \ldots, \begin{pmatrix}X_{t_\ell} \\ \ell/\ell\\ 1\end{pmatrix}, \begin{pmatrix}X_{t_\ell} \\ \ell/\ell\\ 0\end{pmatrix}, \begin{pmatrix}0\\ 0\\ 0\end{pmatrix} \right).
\end{align}
so that the signature of the corresponding piecewise linear path $\tilde{X}_\text{interp}$, contains the terms 
$\phi(\tilde{X}_\text{interp})^{1}=-X_{t_1}$ and $\phi(\tilde{X}_\text{interp}))^{1,2,3}=X_{t_1}-\frac{1}{t_{\ell}-t_1}\int_{t_1}^{t_{\ell}}X_\text{interp}(t)dt$. Therefore, the temporal mean of the piecewise linear interpolation of $X$ can be obtained by a linear combination of two terms of the signature. Furthermore $\phi(\tilde{X}_\text{interp})^{2}=-1/\ell$ hence is a function of the dwell time given by $\Delta t\times(\ell-1)$ where $\Delta t$ is the nanopore sampling period. 

\medskip 
\noindent\textit{Lead-lag.} Another common augmentation is the lead-lag transform~\cite{chevyrev2016primer}, which doubles the dimension of the path in order to capture quadratic variation-type information. Given a discrete signal $X = (X_{t_1}, X_{t_2}, \ldots, X_{t_\ell})$, we define its lead-lag version $\widetilde{X}^{\mathrm{LL}}$ by
\[
\widetilde{X}^{\mathrm{LL}} = \bigl((X_{t_1},X_{t_1}),\ (X_{t_2},X_{t_1}),\ (X_{t_2},X_{t_2}),\ (X_{t_3},X_{t_2}),\ \ldots,\ (X_{t_\ell},X_{t_\ell})\bigr).
\]
In other words, the first coordinate (the ``lead'') is updated one step earlier than the second coordinate (the ``lag''). The piecewise linear interpolation of $\widetilde{X}^{\mathrm{LL}}$ produces a two-dimensional path. The increment, and consequently the signature of the lead-lag path, directly contains information about the consecutive changes in the original signal. This allows the signature to distinguish between, for example, a signal that increases and then decreases versus a signal that just increases. 

These \textit{path transformations} can be combined; for instance, the DENV results use both the invisibility and the lead-lag transforms, and the rRNA and mRNA results implement \Cref{eq:path_aug_1}, which results from applying both the invisibility and time augmentation.

\subsection{Anomaly score design} 
After transforming nanopore current stretches into fixed-dimensional embeddings using the signature transform, we measure the degree of novelty of a signature by computing its nearest neighbor distance \cite{shao2020dimensionless, arrubarrena2024novelty,cass2024variance} in a reference set of signatures of unmodified signals. We whitened the feature vectors using the singular value decomposition (SVD) of the signatures in the reference set. Therefore, our distance metric is data-dependent. 

\medskip 
\noindent\textit{Whitening.} Let $\Phi$ denote the $n\times d$ matrix of the $d$-dimensional signature feature vectors of $n$ reads. That is, each row $\Phi_{i,:}$ stores a signature. We first center the data matrix, by computing the mean over the $n$ samples and removing it to each row. We then compute the singular value decomposition (SVD) of $\Phi$, that is $\Phi=U\Sigma V^\top$ where $U$ is a $n\times n$ orthogonal matrix, $\Sigma$ is a $n\times d$ matrix with the singular values $\sigma_1\geq \sigma_2 \geq \ldots \sigma_d \geq 0$ on its diagonal and all other entries set to zero, and $V$ is a $d\times d$ orthogonal matrix. Denote by $r$ the number of non-zero singular values. The columns of $V$ are orthonormal eigenvectors $v_1, \ldots, v_d$ of $\Phi^\top \Phi$ where $\Phi^\top \Phi v_j=\sigma_i^2 v_j$. We use the SVD to whiten the data so that the result has mean zero and identity covariance matrix. More precisely, the non-zero singular values are scaled using the number of samples $\tilde{\sigma}_j=\sigma_j/\sqrt{n-1}$ and used to rescale each component $\tilde{v}_j=v_j/\tilde{\sigma}_j$. The whitening matrix $W$ is then the $d\times r$ matrix where each column $W_{:,j}$ stores a rescaled component $\tilde{v}_j$. Whitening a data matrix $\Phi$ consists in computing the $n\times r$ matrix $\Phi^\text{whitened}=\Phi W$, so that the new features are less correlated with each other, and all have the same variance. 

\medskip 
\noindent\textit{Data-driven metrics.} Any set $\mathscr{C}$ of \(n\) signatures—feature vectors derived from \(n\) stretches of nanopore signals from IVT transcripts—can be collected into an array \(\Phi\in \mathbb{R}^{n \times d}\). From these signatures, we compute a whitening matrix \(W \in \mathbb{R}^{d \times r}\) that decorrelates and scales the data. Any new signature \(\phi_* \in \mathbb{R}^d\) is transformed into \(\phi_*^\text{whitened} = \phi_* W \in \mathbb{R}^r\). We define an IVT metric as
\begin{align}\label{eq:ivt_norm}
\|\phi\|_\mathscr{C} := \|\phi W\|_2,
\end{align}
which is the Euclidean norm of the transformed signature. Each initial collection of \(n\) stretches of IVT nanopore signals yields a different metric. This metric is invariant to linear rescaling of signature features and changes in measurement units. To analyze a transcript of interest, we collect IVT reads corresponding to its canonical nucleotide sequence. After alignment, we compute feature vectors from the nanopore signal stretches for each potition and derive the corresponding whitening matrices. This yields a different metric for each site. 

\medskip 
\noindent\textit{Nearest neighbor distance.} We use the metric to compare each individual feature vector from the new sequence with equivalent vectors from the corpus of IVT unmodified feature data. This distance may be described using the Mahalanobis distance to each element in the corpus. Given the signature $\Signature{X}$ of the streamed measurements collected while a $k$-mer $b$ is in the sensing region at the $i^{th}$ site in the sequence, we define the anomaly score $\score{\Signature{X}}{\mathscr{C}_i}$ with respect to a set $\mathscr{C}_i$ of signatures as the nearest neighbor distance, 
\begin{align}
    \score{\Signature{X}}{\mathscr{C}_i}=\min_{\Signature{Y}\in \mathscr{C}_i}\|\Signature{X} -\Signature{Y}\|_{\mathscr{C}_i}
\end{align}
where $\|\cdot\|_{\mathscr{C}_i}$ is the norm defined in \Cref{eq:ivt_norm}.

\subsection{Conformal p-values and statistical inference}
\label{ssec:conformal_p_values}

\noindent\textit{Split conformal p-values.} We use a calibration dataset to gauge how concentrated the feature vectors of the IVT data are, relative to the distance and set the thresholds between normal and anomalous feature vectors. Specifically, we compute conformal p-values\cite{bates2023testing} to decide how unusual or exceptional the score is for the input data stream. This is a well-established statistical procedure for non-parametric testing for outliers. Consider a \emph{fixed site}. To compute so-called \textit{split conformal p-values}, the calibration set for the site is separated from the IVT data prior to the metric in \Cref{eq:ivt_norm} being calculated. 
More precisely, given a set of $n$ signatures $\Signature{X^{(1)}}, \ldots, \Signature{X^{(n)}}$ of modification-free $k$-mer signals, we split it into two sets $\mathcal{D}_{\text{ivt}}=\mathcal{D}_{\text{train}}\cup \mathcal{D}_{\text{cal}}$ and only use $\mathcal{D}_{\text{train}}$ to build the nearest neighbor score. The (split) conformal p-value of a new path $X$ is then defined by
\begin{align}\label{eq:split_conformal_p_value}
   \hat u(X;\mathcal{D}_\text{cal})
\;=\;
\frac{1 + \bigl|\{i\in \mathcal{D}_\text{cal} : s(\Signature{X^{(i)}}) \geq s(\Signature{X})\}\bigr|}{1+|\mathcal{D}_\text{cal}| } 
\end{align}
where we omit the corpus dependency in the score to simplify notation.
We note that there are several types of other conformal p-values such as \textit{calibration-conditional} \cite{bates2023testing}, \textit{full} \cite{lee2025full} and \textit{leave-one-out}, \textit{cross} and \textit{bootstrap} \cite{hennhofer2024leave} conformal p-values. The split conformal p-values can then be corrected using, e.g., the \textit{Benjamini-Hochberg (BH)} procedure \cite{bates2023testing} and Storey's correction \cite{storey2004strong} to control the false discovery rate. However, due to the dependency on the calibration set, these p-values are not always valid in a multiple testing setting. For example, the Fisher's combination test becomes invalid. They need to be adjusted before being combined. The resulting adjusted p-values are referred to as \emph{calibration-conditional conformal p-values}\cite{bates2023testing}.

\noindent\textit{Full conformal p-values.} When the coverage in the IVT sample is low, after using most of the reads to compute the nearest neighbor metric, there might not be enough reads left to construct the calibration set. In this setting, it is desirable to make use of the full reference IVT sample $\mathcal{D}_{\text{ivt}}$ and take $\mathcal{D}_{\text{train}}=\mathcal{D}_{\text{ivt}}$ to   compute the \gls{nns} scores and $\mathcal{D}_{\text{cal}}=\mathcal{D}_{\text{ivt}}$ to compute the p-values by \Cref{eq:split_conformal_p_value} to conduct the inference. In this case, these are called \textit{full} conformal p-values. They are defined as in \Cref{eq:split_conformal_p_value}, except that the score is now given by the second nearest neighbor distances in $\mathcal{D}_{\text{ivt}}$.They are still marginally superuniform, but some multiple testing procedure may be invalid.

\medskip 

\noindent\textit{Remark.} There are limitations in using the two sample KS test for calling modified sites. This test compares the distributions of anomaly scores between an \acrshort{ivt} sample and the experimental sample using the test statistic $T := \sup_x \bigl|F_{n}(x) - G_{m}(x)\bigr|$ where $F_{n}$ and $G_{m}$ are the empirical counterparts of $F,G$, the distributions of scores from the two samples. When only a small fraction of reads are genuinely anomalous, it may lack power to detect a shift. Under a contaminated model $G=(1-\epsilon)F+\epsilon\tilde{G}$, with contamination ratio $\epsilon \approx 0.05$, even for relatively large samples ($m=n=500$), the KS test has low sensitivity when $F$ and $\tilde{G}$ are both Gaussian with variance $1$ and $100$ respectively \cite{makarov2016some}. Additionally, this test lacks power when the number of reads $m$ in the experimental sample is small, limiting its applicability to the setting where \acrshort{ivt} and experimental samples are both large and modification levels are high.

\subsection{Multiple testing correction}\label{ssec:conformal_inference}

\noindent\textit{Single molecule multiple testing problem.}
Our method simultaneously tests the hypothesis that the nanopore signal is unmodified across many sites and reads, creating a multiple testing problem that requires statistical correction. More formally, we test $M$ null hypotheses $H_{i,j}:X^{(i,j)}\sim \mu^{j}$ where, $X^{(i,j)}$ is the nanopore stretch of the $i^{th}$ read at the $j^{th}$ site, and $\mu^j$ is the distribution, on path space, of the unmodified nanopore signal at the $j^{th}$ site. Without correction, testing thousands of read-site pairs would generate many false positives by chance alone. Since the conformal p-values are super-uniform, the average number of false rejections is $\mathbb E[\sum_{i,j}\mathds 1[\hat{u}(X^{(i,j)};\mathcal{D}^j_\text{cal})\leq \alpha]\leq M\alpha$, where the average is taken over the $X^{(i,j)}$ and the calibration datasets $\mathcal{D}^j_\text{cal}$ used to compute the conformal p-values. This number may be very large, and we need to correct the set of critical values $(\alpha_{i,j})_{i,j=1}^{m}$. We apply the \gls{bh} procedure (possibly with Storey's correction) to control the false discovery rate. Specifically, our algorithm returns a set $S$ of site-read pairs seen as outliers. Denoting by $\mathcal{H}_0$ the set of inliers (i.e. $\mathcal{H}_0=\{(i,j)\in[M]: H_{i,j} \text{ is true}\}$) and by $\mathcal{H}_1$ the set of outliers (i.e. $\mathcal{H}_1=\{(i,j)\in[M]: H_{i,j} \text{ is false}\}$), the false discovery rate is given by
\begin{align}
\mathrm{FDR} := \mathbb{E}[\mathrm{FDP}], 
\quad\text{where}\quad
\mathrm{FDP} 
:= \frac{|S\cap \mathcal{H}_0|}{\max\{|S|,1\}}.
\end{align}
Here, a FDR of $5\%$ means that among all read-site pairs called significant (anomalous), $5\%$ of these are truly null (non-anomalous) on average over the $X^{(i,j)}$ and the $\mathcal{D}^j_{\text{cal}}$. To control the FDR at level $q$, the \acrshort{bh} procedure consists in sorting the p-values $p_{(1)}\leq \ldots \leq p_{(M)}$, and finding the largest $r$ such that $p_{(r)}\leq qr/M$. All hypotheses below the threshold $qr/M$ are then rejected. In practice, we apply FDR control independently at each genomic position, correcting p-values across all reads covering that site. We optionally use Storey's correction, which estimates the proportion $\hat{\pi}_0$ of unmodified reads at each site to increase detection power.

\medskip 

\noindent\textit{Site-level multiple testing problem.} We also test $N$ global nulls $H_j: X^{(1,j)}, \ldots, X^{(n_j,j)}\sim \mu^j$, where each test asks whether the site $j$ contains any outlying read. It is less informative than the previous problem, that consists in identifying the outlying reads, but it is of interest for transcriptome-wide analyses. We use the Fisher's combination test after calibration-conditional adjustment of the conformal p-values.  As we conduct this test simultaneously over $N$ sites, we use the BH procedure. 

\subsection{Testing for differential anomaly rate between samples}
We define an anomaly rate at a site by counting the number of conformal p-values below a chosen threshold $\alpha$, divided by the number of reads $n$ in that sample. Formally, the anomaly rate is the random variable
    \begin{align}\label{eq:emp_fpr}
        Y=\hat{F}_n\left(\hat{F}^{-1}_m\left(\frac{\alpha (m+1)}{m}\right)\right)
    \end{align}
    where $\hat{F}_n$ and $\hat{F}_m$ denote the empirical distribution of the \acrshort{nns} based on $n$ native reads and $m$ calibration reads, respectively. Using classical results for the order statistics of uniform variables, the distribution of $Y$ is obtained in closed form
    \begin{align}\label{eq:beta_binom}
        n\hat{F}_n(\hat{F}^{-1}_m(p))\sim \text{Beta-Binomial}(n, \alpha =\lceil mp\rceil, \beta=m-\lceil mp\rceil+1),
    \end{align}
    Thus, the number of anomalies in a sample follows a Beta–Binomial distribution, with parameters determined by the calibration set size $m$ and the chosen threshold $\alpha$. To compare anomaly rates between two conditions, we use a parametric bootstrap test for two proportions under the Beta–Binomial model. This accounts both for finite calibration uncertainty (through $m$) and for variability in the native sample size $n$. These numbers are all reported in our BED file which allows us to do the test when analyzing the results. In the paper, we use this statistical test to identify m6A modifications using a METTL3 knockout, but this test could also enable differential analysis across conditions \cite{pratanwanichIdentificationDifferentialRNA2021, hewel2024direct}.

\subsection{Comparisons}
Direct comparison of our method with existing nanopore-based modification callers is not straightforward, as the tools rely on fundamentally different principles. Classification-based methods are trained to recognize a fixed set of modification types based on labeled current signal data, and their performance depends critically on the composition of the training set. In contrast, our approach is reference-based: it evaluates conformity of ionic current signals to unmodified $k$-mer distributions, without requiring prior knowledge of modification types. As a result, any apparent differences in performance could reflect not only sensitivity to modifications but also preprocessing choices, such as our requirement for at least 1,000 in vitro transcribed (IVT) reads per site, or the need in other tools for specific control samples. Moreover, a rigorous benchmark for comparison would require a balanced database spanning a wide diversity of RNA modifications. Existing resources, such as RMBase, are heavily skewed toward a subset of well-studied modifications (notably m6A), whereas our method is designed to highlight a broader spectrum of potential modifications, including those that have not yet been systematically characterized.

\subsection*{Data availability}
For the \textit{E. coli} rRNA and the mouse mRNA analyses, we used publicly available datasets from the NCBI BioProject database with project ID PRJNA634693~\cite{Val_biorxivpreprint} (Direct RNA nanopore sequencing kit: SQK-RNA002, MinION flow cell:FLO-MIN106D) and PRJNA1412945 (Direct RNA nanopore sequencing kit: SQK-RNA004, PromethION flow cell:FLO-PRO004RA) respectively. 
The dengue sequencing data (gRNA and sfRNA) will be uploaded to the NCBI SRA archive, while the oligonucleotide data are available in Chia Ching Wu et al~\cite{Wu2025.03.17.643699}.

\subsection*{Code availability}
To protect the commercial potential of this work, the code is currently not publicly released. 

\subsection*{Acknowledgements} 
This work was supported by the UK Engineering and Physical Sciences Research Council (EPSRC) [EP/S026347/1 to M.L., P.A., T.L., and T.C.]; The Alan Turing Institute under the EPSRC grant [EP/N510129/1 to M.L., P.A., T.L., and T.C.]; the Data Centric Engineering Programme under the Lloyd’s Register Foundation grant [G0095 to T.L.]); the Defence and Security Programme funded by the UK Government and the Office for National Statistics \& The Alan Turing Institute [strategic partnership to T.L.]; the Hong Kong Innovation and Technology Commission [InnoHK Project CIMDA to T.L.]; The Royal Society International Exchanges grant "SigMod" [IES/R2/232269 to J.B., A.P., F.N.P. and S.d.G.]; the Deutsche Forschungsgemeinschaft (DFG, German Research Foundation) [RTG2727–445549683 to V.G.; SVB TRR319–RMaP to S.d.G. and V.G.; TRR319-RMaP-439669440 A05 to A.R. and A04 to F.N.P.]. We thank all members of the collaborating teams at the University of Oxford, Imperial College London, the German Cancer Research Center (DKFZ), and the University of Warwick for their support and discussions throughout the development of this work. The authors acknowledge the administrative and computational support provided by the Mathematical Institute at Oxford and the Department of Mathematics at Imperial College London.

\subsection*{Author contributions}
M.L. conducted the initial analyses establishing the power of signature-based features for RNA modification detection. She designed and developed the statistical and computational framework, including read-level statistics, p-value aggregation, and multiple-testing control, and implemented the core software supporting the analyses in the study and the manuscript visualizations. She performed the E. coli and mouse mRNA analyses, establishing the method performance and identifying candidate sites, and provided analytic guidance on the Dengue dataset. M.L. structured and drafted the manuscript, integrating feedback from co-authors. P.A. developed and maintained the data-processing and analysis pipelines for the E. coli, Dengue, and mRNA datasets she worked on, building on the shared statistical framework. She conducted exploratory computational studies on the E. coli datasets and carried out the analyses of the Dengue virus datasets that detected and contextualised the novel sfRNA modification using this framework. She contributed figures, dataset management, and cross-disciplinary communication throughout the project, and co-wrote the Dengue results section with S.D.G. S.D.G. generated and curated the Dengue RNA and synthetic datasets; designed and supervised the orthogonal validation experiments; established the preprocessing procedures for the raw signal data; and contributed the biological interpretation of both the Dengue and sfRNA results. J.B. provided statistical guidance and contributed to validation methodology. T.C. contributed to statistical methodology, conformal inference design, and overall supervision of the mathematical components. V.G. created the raw cell lines IVTs. I.S.N.d.V. generated all dengue data (IVT and sfRNA). A.P. initiated the collaboration and advised on experimental framing and data interpretation. I.T. worked with S.D.G and M.L. on early versions of the data pipeline. A.R. and W.C.C. performed the validation of the findings on viral sfRNA. F.N.P. supervised the biological and experimental components, oversaw validation and data generation, coordinated cross-site collaboration, and guided integration across disciplines. T.L. conceived and supervised the overall project, integrating the mathematical and biological components and leading the model-free anomaly detection strategy. All authors reviewed and approved the final manuscript.

\subsection*{Additional information} 
\textbf{Open access} For the purpose of open access, the authors have applied a Creative Commons Attribution (CC BY) license to any Accepted Manuscript version arising.

\bibliographystyle{naturemag}
\bibliography{bibliography.bib}

\clearpage

\appendix
\section{Extended Data}

\begin{figure}[h]
\centering
   \begin{subfigure}{0.90\textwidth}
\includegraphics[width=1.\linewidth]{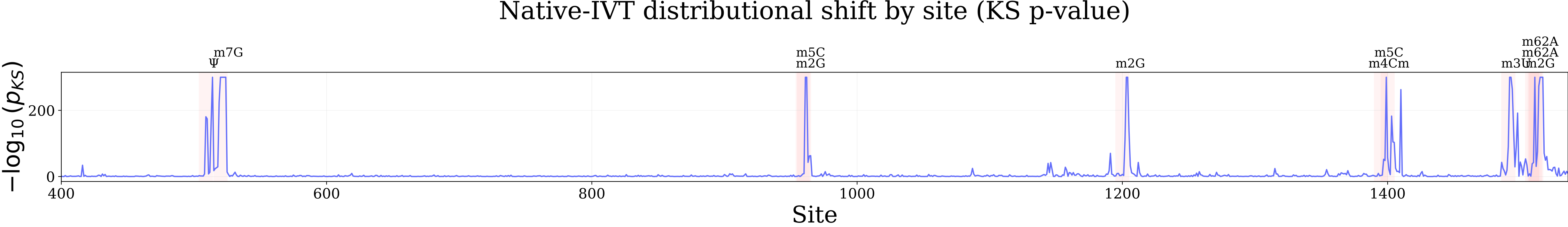}
\caption{}
\label{fig:ecoli16s_ks}
    \end{subfigure}
\\
   \begin{subfigure}{0.90\textwidth}
\includegraphics[width=1.\linewidth]{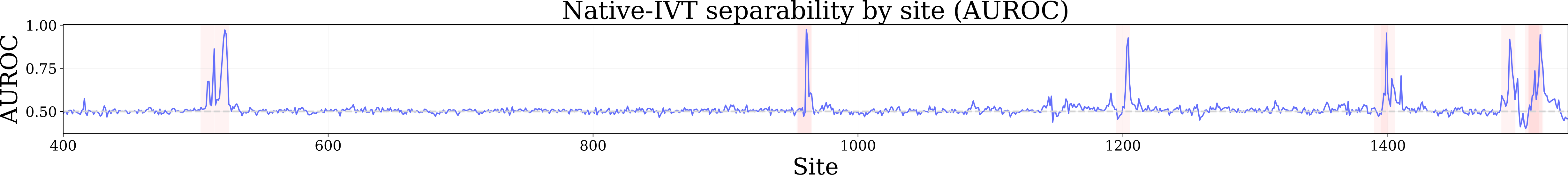}
\caption{}
\label{fig:ecoli16s_auroc}
    \end{subfigure}
\\
   \begin{subfigure}{0.90\textwidth}
\includegraphics[width=1.\linewidth]{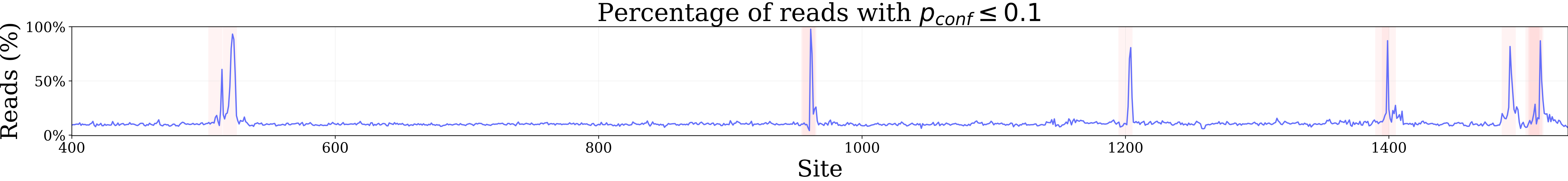}
\caption{}
\label{fig:ecoli16s_anomaly_rate}
    \end{subfigure}
\\
   \begin{subfigure}{0.90\textwidth}
\includegraphics[width=1.\linewidth]{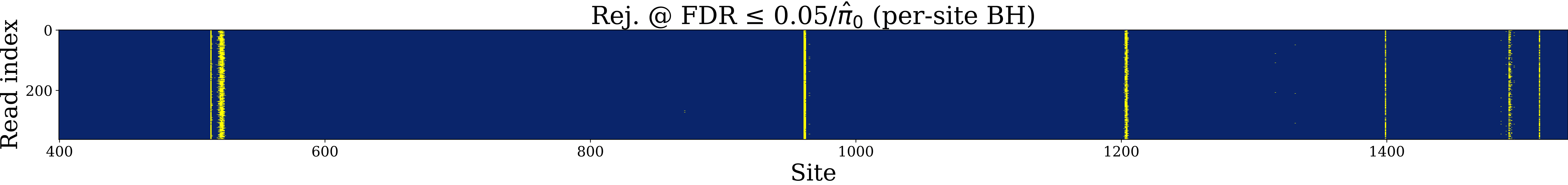}
\caption{}
\label{fig:ecoli16s_persite_perread_decisions}
    \end{subfigure}
\\ 
  \begin{subfigure}{0.90\textwidth}
\includegraphics[width=1.\linewidth]{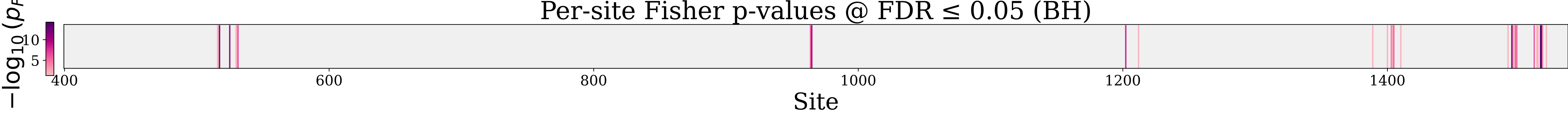}
\caption{}
\label{fig:ecoli16s_persite_decisions}
    \end{subfigure}
    \caption{\textbf{Evaluation on ribosomal RNA modifications in E. coli 16S.} (a) KS test p-values comparing the score distributions between native and \acrshort{ivt} reads for each site. (b) AUROC values quantifying the performance of the anomaly detector. (c) Percentage of individual reads with
a score exceeding the 0.90 quantile of the calibration scores. (d) Single-molecule (conformal) p-values and FDR control. At a site, the conformal p-values are thresholded at the BH threshold. (e) Per-site (Fisher) combination test with FDR control. Conditional-calibrated split-conformal p-values (asymptotic, $\delta=0.01$) are combined per site using Fisher’s method and corrected by \gls{bh} with Storey at level $0.05$. The heatmap shows the resulting p-values (light gray values are non-significant).}
\label{fig:ecoli16s}
\end{figure}

\begin{figure}[h]
    \centering
    \includegraphics[width=0.9\linewidth]{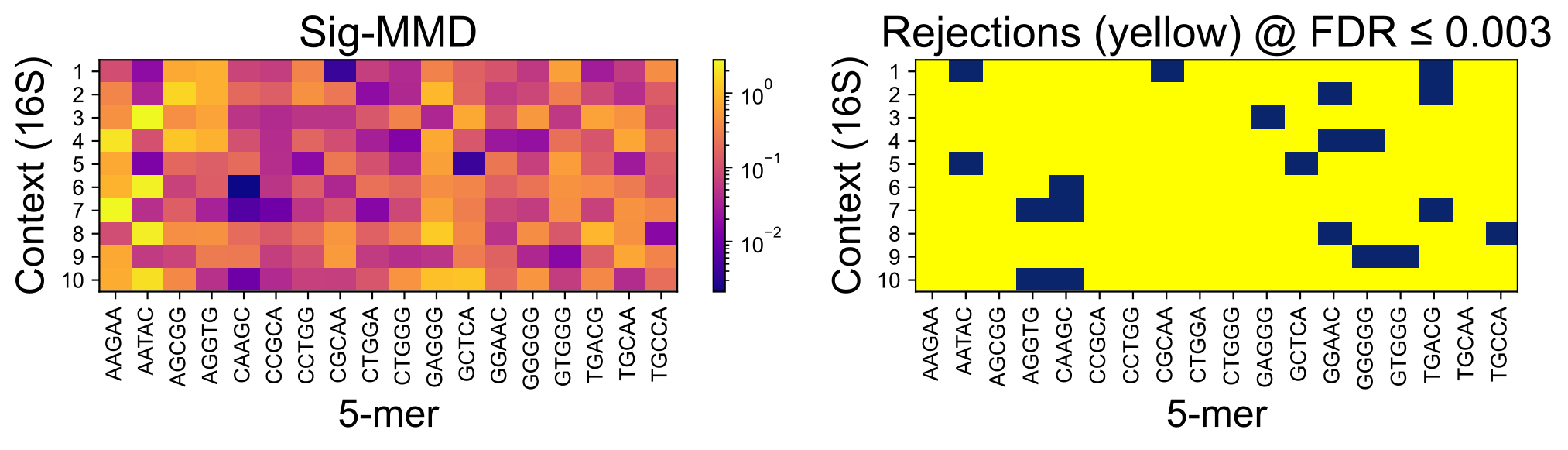}
    \caption{\textbf{Variability of 5-mer signal distributions with the context.} Permutation-based two sample Sig-MMD test\cite{chevyrev2022signature} performed with $1000$ permutations. The 5-mer were selected by taking all 5-mers in E. coli 16S that appear in exactly 5 different positions. The MMD test is conducted on each of the 10 ordered pairs of positions, for each 5-mer, with 1000 sample paths (reads).}
    \label{fig:MMD_test}
\end{figure}

\begin{figure}[H]
    \centering
    
    \begin{subfigure}{0.49\linewidth}
        \centering
        \begin{minipage}{0.45\linewidth}
            \centering
            \textbf{Modified}
        \end{minipage}%
        \begin{minipage}{0.45\linewidth}
            \centering
            \textbf{Unmodified}
        \end{minipage}
        
        \includegraphics[width=0.45\linewidth]{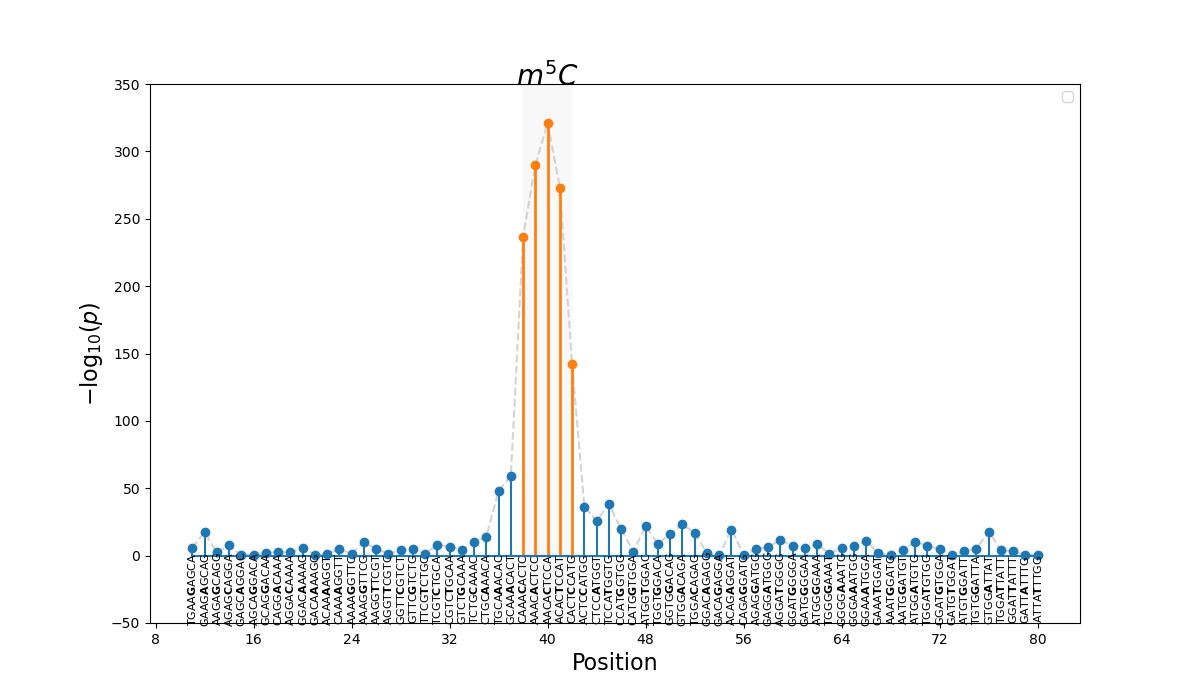}
        \includegraphics[width=0.45\linewidth]{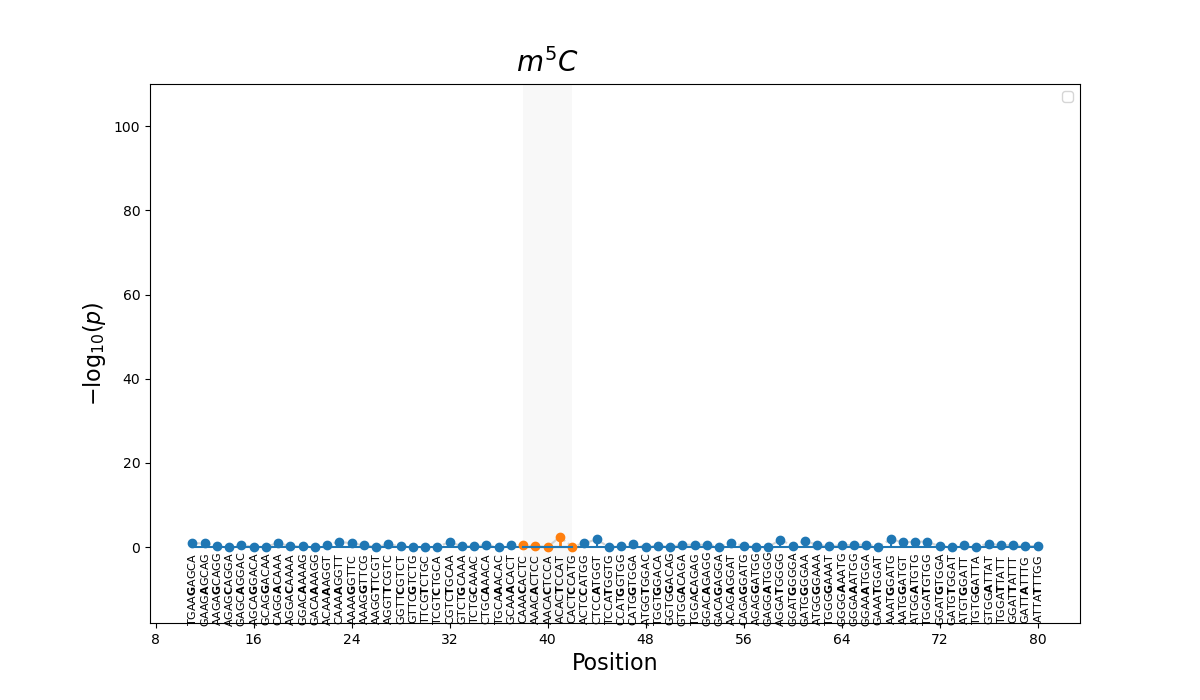}\\
        
        \includegraphics[width=0.45\linewidth]{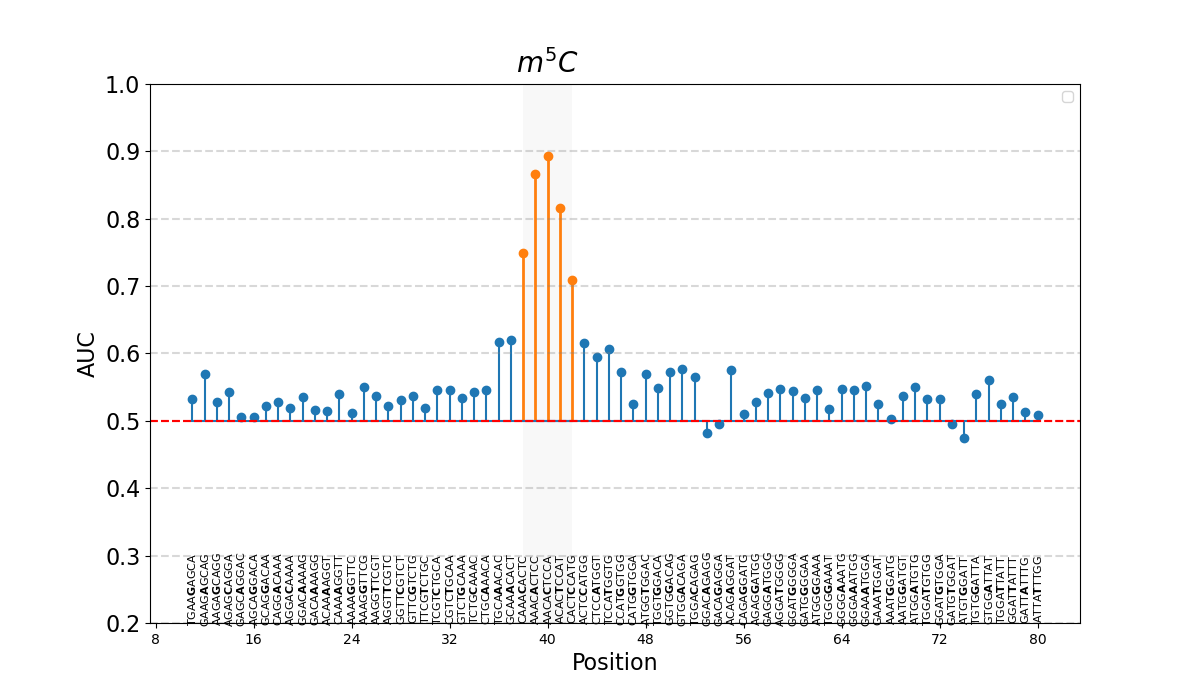}
        \includegraphics[width=0.45\linewidth]{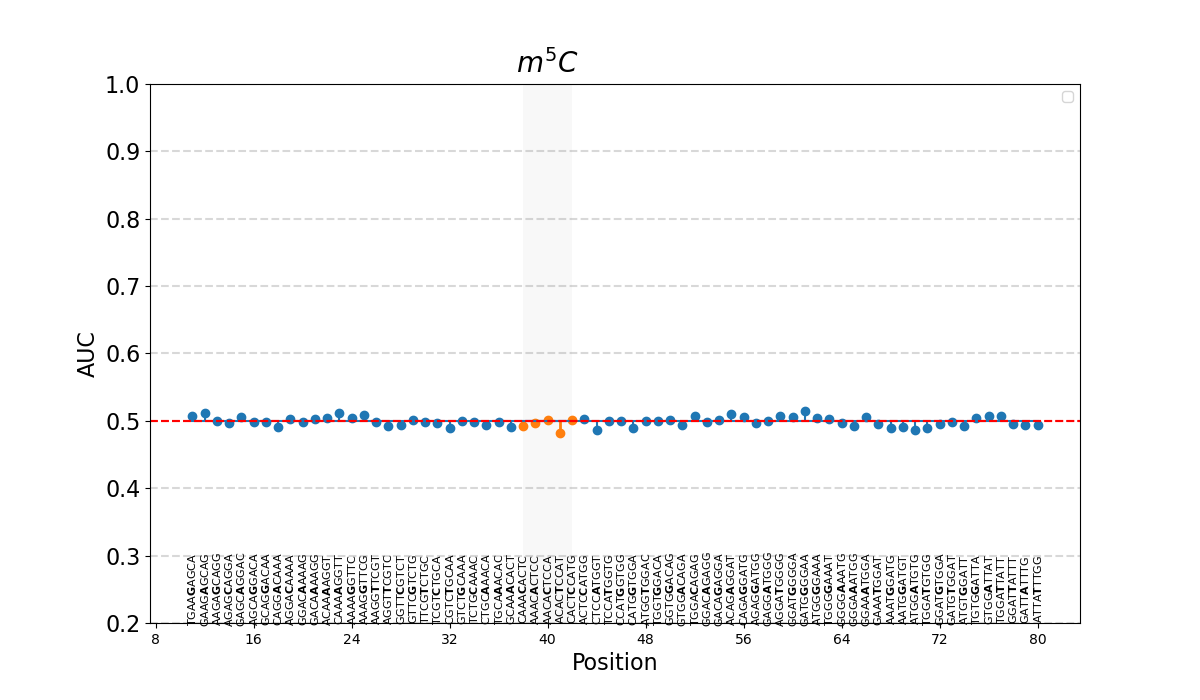}
        
        \caption{}
        \label{fig:DENVOligoLowRegionKSAUC}
    \end{subfigure}
    \hfill
    \begin{subfigure}{0.49\linewidth}
        \centering
        \includegraphics[width=\linewidth]{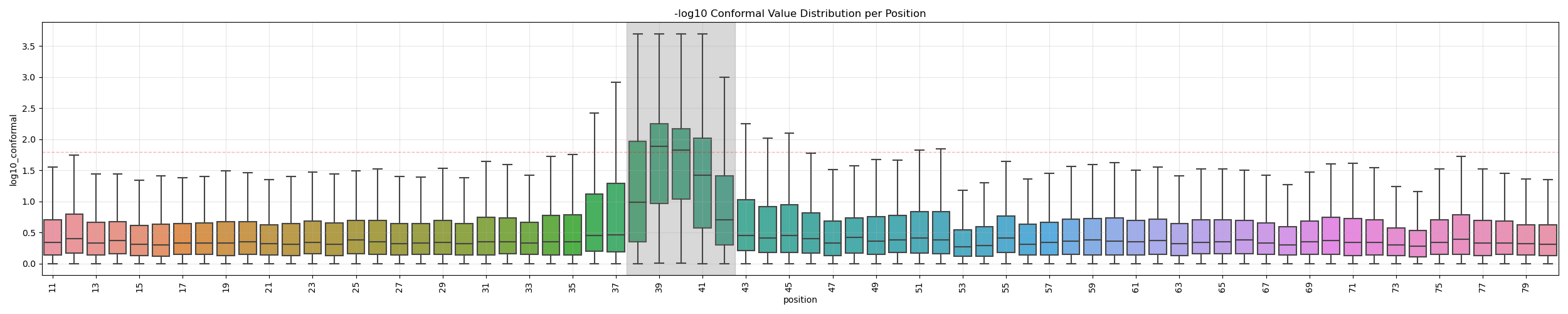}
        \caption{}
        \label{fig:OligoConformalThresh}
    \end{subfigure}
\caption{(a) \textbf{DENV m5C-modified oligonucleotides.} \textit{Top left:} KS test $p$-values for m5C-modified. 
        \textit{Top right:} KS test $p$-values for unmodified. 
        \textit{Bottom left:} AUC values for m5C-modified. 
        \textit{Bottom right:} AUC values for unmodified. (b) \textbf{Conformal $p$-values for DENV m5C-modified oligo.} Around position 40.}
\end{figure}

\begin{figure}[H]
\includegraphics[width=1.\linewidth]{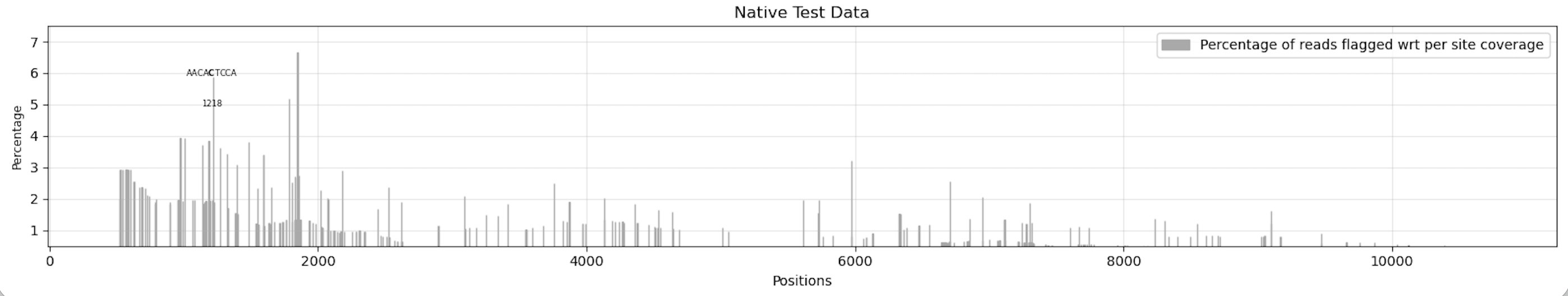}
\caption{Percentage of individual reads thresholded at per-site BH level all positions in the DENV gRNA. The plot shows results using native data, which may contain modifications. Positions 1218, are above 5\%.}
\label{fig:DENVgRNAIVTandTEST}
\end{figure}

\begin{figure}[h]
    \centering

\includegraphics[width=0.3\linewidth]{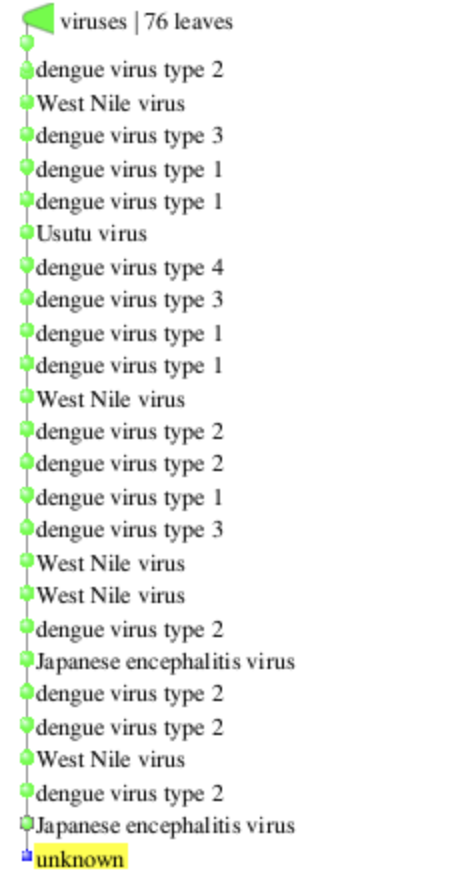}
    \caption{High modification sequence distance tree.}
    \label{fig:Distance tree - Blast H m region}
    \end{figure}
\end{document}